\documentclass[12pt]{article}
\usepackage{graphicx}
\usepackage{color}
\begin{document}
% =================================================
% \input{title.tex}
% ===============================================================
% File title.tex .
% Last modified: 30 August 2023
% =========================================================
\begin{center}
\textbf{Test of the Orbital-Based LI3 Index as a Predictor
of the Height of the $^3$MLCT $\rightarrow$ $^3$MC
Transition-State Barrier for Gas-Phase [Ru(N$^\wedge$N)$_3$]$^{2+}$ 
Polypyridine Complexes}
\end{center}
\normalsize

\vspace{0.5cm}

\noindent
Denis Magero\\
{\em School of Science, Technology and Engineering, Department of Chemistry and Biochemistry,
Alupe University, P.O.\ Box 845-50400, Busia, Kenya\\
e-mail: dmagero@au.ac.ke}

\vspace{0.5cm}

\noindent 
Ala~Aldin M.\ H.\ M.\ Darghouth \\
{\em College of Sciences, University of Mosul, Al Majmoaa Street, Mosul, 
41002 Iraq.\\
% Iraq Zip Code: 41002.\\
e-mail: aladarghouth@uomosul.edu.iq}

\vspace{0.5cm}

\noindent
Mark E.\ Casida\\
{\em Laboratoire de Spectrom\'etrie, Interactions et Chimie th\'eorique 
(SITh),
D\'epartement de Chimie Mol\'eculaire (DCM, UMR CNRS/UGA 5250),
Institut de Chimie Mol\'eculaire de Grenoble (ICMG, FR-2607), 
Universit\'e Grenoble Alpes (UGA)
301 rue de la Chimie, CS40700, F-38058 Grenoble Cedex 9, France\\
e-mail: mark.casida@univ-grenoble-alpes.fr}

% \vspace{0.5cm}

% \noindent
% Date of Publication: \today \hspace{0.25cm} (MS 6.10)

\vspace{0.5cm}

\begin{center}
{\bf Abstract}
\end{center}

Ruthenium(II) polypyridine compounds often luminesce but the luminescence 
lifetime depends upon the precise nature of the ligands.  This luminescence
lifetime is thought to be controlled by the barrier to conversion from an
initial phosphorescent triplet metal-ligand charge transfer ($^3$MLCT) 
state to a nonluminscent triplet metal-centered ($^3$MC) state which decays
nonradiatively.  Earlier work [{\em J.\ Photochem.\ Photobiol.\ A} 
{\bf 348}, 305 (2017)] took room temperature and liquid nitrogen (77 K) 
lifetimes from a large previously-published database [{\em Coord.\ Chem.\ Rev.}
{\bf 84}, 85 (1988)] and extracted empirical average $^3$MLCT $\rightarrow$ 
$^3$MC transition state (TS) barrier heights ($E_{ave}$s) which were {\em not}
believed to be quantitative but which were believed to capture the 
trends in the true barrier heights correctly.  These were then used 
together with information from partial density of states calculations 
[{\em J.\ Photochem.\ Photobiol.\ A} {\bf 276}, 8 (2014)] to derive 
several orbital-based luminescence indices of which the
third (LI3) was based upon frontier-molecular-orbital-like
ideas and correlated linearly with values of $E_{ave}$.  As it is known
that $E_{ave}$ is a large underestimate of the true $^3$MLCT $\rightarrow$ 
$^3$MC TS barrier height in the case of the {\em tris}bipyridine
ruthenium(II) cation $\{$ [Ru(bpy)$_3$]$^{2+}$ $\}$, but accurate TS 
barrier heights are difficult to obtain experimentally, it was judged useful 
to verify the ideas
used to derive the LI3 index by calculating the energetics of the gas-phase
$^{3}$MLCT $\rightarrow$ $^{3}$MC reaction for a series of ruthenium(II) 
{\em tris} bipyridine complexes using the same density functional and
basis sets used in calculating LI3.  
Specifically, four closely-related bipyradine complexes 
$\{$ [Ru(N$^\wedge$N)$_3$]$^{2+}$ with N$^\wedge$N = bpy ({\bf 6}), 
4,4'-dm-bpy ({\bf 70}), 4,4'-dph-bpy ({\bf 73}), and 4,4'-DTB-bpy ({\bf 74}) 
$\}$ were used for these calculations.
We examine the {\em trans} dissociation mechanism in great detail
at the B3LYP/6-31G+LANLDZ(Ru) level of detail and uncover a two part
mechanism.  In the first part, the electron is transferred to a single
ligand rather than symmetrically to all three ligands.  It is the two
Ru-N bonds to this ligand which are equally elongated in the transition
state.  The intrinsic reaction coordinate then continues down a ridge
in hyperspace and bifurcates into one of two symmetry-equivalent $^3$MC 
structures with elongated {\em trans} bonds.  Interestingly, no significant
difference is found for the TS barriers for the four complexes treated
here.  Instead, LI3 is linearly correlated with the energy difference
$\Delta E = E(\mbox{$^3$MLCT}) - E(\mbox{$^3$MC})$.
While this work shows that LI3 predicts the total energy difference and this does correlate well with the luminescence lifetime, we do not have a detailed understanding yet of how this happens and are also wary of oversimplicity as there are likely other excited-state reactions occurring on similar time scales which may also impact the luminescence lifetimes.

\vspace{0.5cm}
 \begin{center}
 \includegraphics[width=0.8\textwidth]{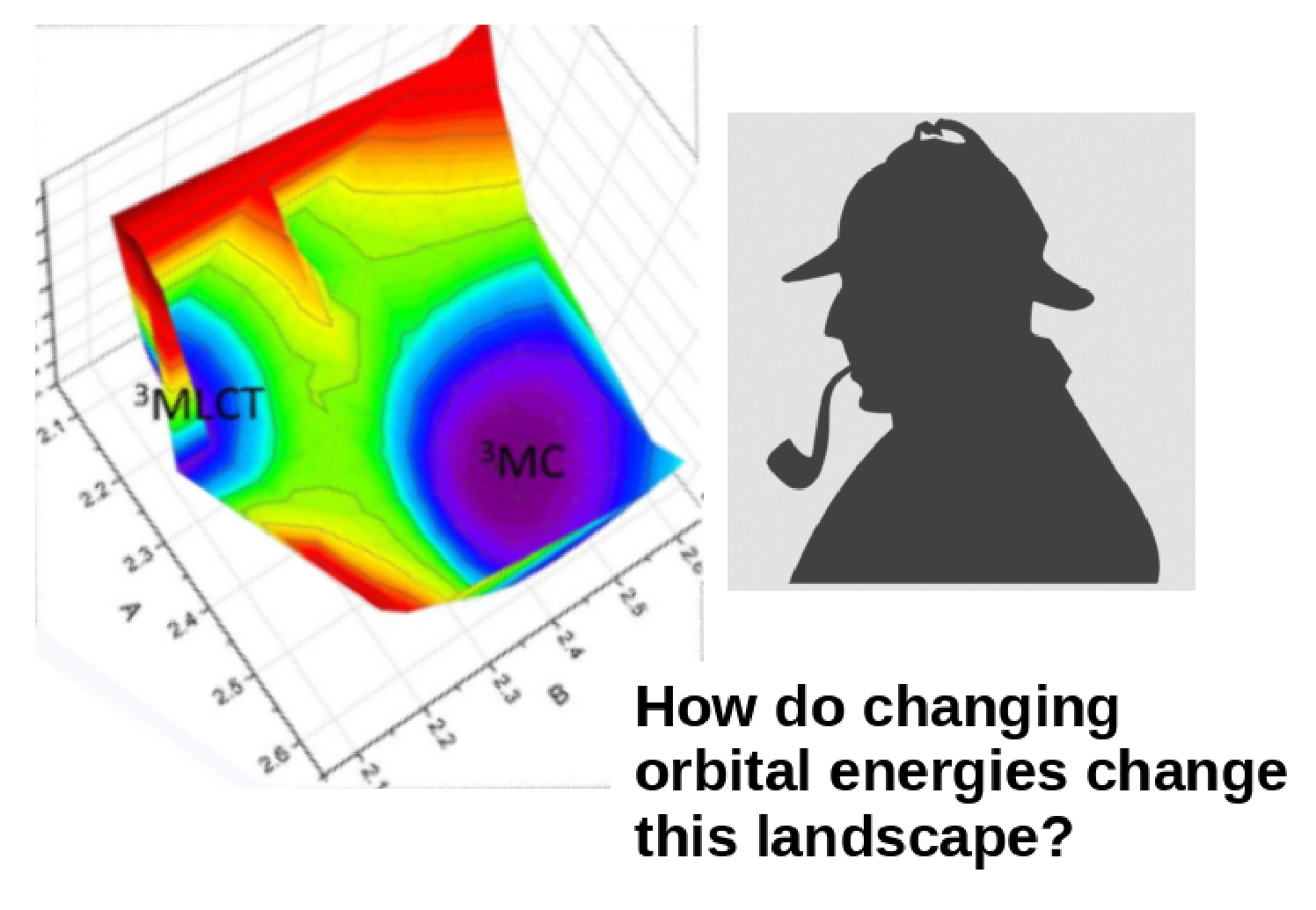}\\
 {\bf Graphical Abstract}
 \end{center} 

\vspace{0.5cm}

% % -------------------------------------------------
% {\color{green}
% \tableofcontents
% }
% \newpage
% \input{notes.tex}
%%%%%%%
% EOF %
%%%%%%%
% ----------------------------------------------------------
\section{Introduction}
\label{sec:intro}
% \input{intro.tex}
% ========================================
% file: intro.tex
% last updated: 30 August 2023
% ========================================

\begin{figure}
   \includegraphics[scale=1.5]{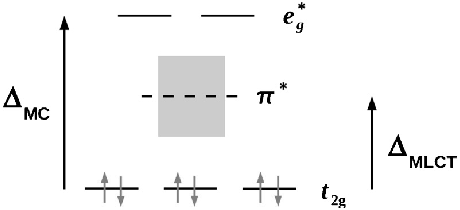}
   \centering
   \caption{
   Pseudo-octahedral ligand field theory diagram for ruthenium(II) complexes. 
   \label{fig:pseudolft}
   }
\end{figure}

Luminescent ruthenium(II) complexes elicit immense interest 
\cite{MMAC20,SCC+94,N82,LKW99,BJ01,DTL+03,HKZ03,MH05,%
NSF+08, JWW+10, SAH+20,BCCV21} owing to their wide range of applications,
including photochemical molecular devices, % (PMDs), 
biological sensors, organic light emitting diodes, % (OLEDS), 
and biomedical applications, 
% -----------------------------
\marginpar{\color{blue} LFT}
% ----------------------------
among others.  The commonly assumed ligand-field theory (LFT) 
mechanism explaining this
luminescence begins with the $d^6$ ruthenium complex initially in
% -----------------------------
\marginpar{\color{blue} GS}
% ----------------------------
its closed-shell ground-state ($^1$GS) configuration, 
\begin{equation}
  \mbox{metal: } \begin{array}{c}
  \underbrace{[\,\,\, , \,\,\,] [\,\,\, , \,\,\, ] }_{e_g^*} \\
  \underbrace{[\uparrow, \downarrow] [\uparrow, \downarrow]
  [\uparrow,\downarrow]}_{t_{2g}} 
  \end{array} + \mbox{ligands: } 
  \underbrace{[\,\,\, , \,\,\,] [\,\,\, , \,\,\,] \cdots}_{\pi^*} \, .
  \label{eq:intro.1}
\end{equation}
(The $^1$X and $^3$X pre-exponents denote respectively that the state X 
% -----------------------------
\marginpar{\color{blue} $^1$X , $^3$X}
% ----------------------------
is a singlet or triplet.  Note that the $t_{2g}$ orbital is nonbonding 
but that the $e_g^*$ orbitals are antibonding.  We have also indicated 
some unoccupied $\pi^*$ orbitals on the ligands.)  
{\bf Figure~\ref{fig:pseudolft}} emphasizes that these
$\pi^*$ ligand orbitals are within the LFT gap of the metal.
% -----------------------------------------------------------
% -----------------------------------------------------------
Upon absorption of a photon, an electron is excited
from the metal to the ligand to create a high-lying singlet 
% -----------------------------
\marginpar{\color{blue} MLCT}
% ----------------------------
metal-ligand-charge-transfer ($^1$MLCT) configuration, 
\begin{equation}
  \mbox{metal: } \begin{array}{c}
  \underbrace{[\,\,\, , \,\,\,] [\,\,\, , \,\,\, ] }_{e_g^*} \\
  \underbrace{[\uparrow, \downarrow] [\uparrow, \downarrow]
  [\downarrow,\,\,\, ]}_{t_{2g}} 
  \end{array} + \mbox{ligands: } \underbrace{[\uparrow , \,\,\,] [\,\,\, , \,\,\,] \cdots}_{\pi^*} \, ,
  \label{eq:intro.2}
\end{equation}
which relaxes by non-radiative relaxation to the lowest $^3$MLCT state,
\begin{equation}
  \mbox{metal: } \begin{array}{c}
  \underbrace{[\,\,\, , \,\,\,] [\,\,\, , \,\,\, ] }_{e_g^*} \\
  \underbrace{[\uparrow, \downarrow] [\uparrow, \downarrow]
  [\uparrow,\,\,\, ]}_{t_{2g}} 
  \end{array} + \mbox{ligands: } \underbrace{[\uparrow , \,\,\,] [\,\,\, , \,\,\,] \cdots}_{\pi^*} \, .
  \label{eq:intro.3}
\end{equation}
its closed-shell ground-state ($^1$GS) configuration, 
\begin{equation}
  \mbox{metal: } \begin{array}{c}
  \underbrace{[\,\,\, , \,\,\,] [\,\,\, , \,\,\, ] }_{e_g^*} \\
  \underbrace{[\uparrow, \downarrow] [\uparrow, \downarrow]
  [\uparrow,\downarrow]}_{t_{2g}} 
  \end{array} + \mbox{ligands: } 
  \underbrace{[\,\,\, , \,\,\,] [\,\,\, , \,\,\,] \cdots}_{\pi^*} \, .
  \label{eq:intro.4}
\end{equation} 
% ----------------------------
Energetically close to this state is a triplet metal-centered ($^3$MC) 
state which mixes with the $^3$MLCT state to form an avoided crossing,
allowing repopulation of the $e_g^*$ orbitals via back transfer of an
electron,
\begin{equation}
  \mbox{metal: } \begin{array}{c}
  \underbrace{[\uparrow , \,\,\,] [\,\,\, , \,\,\, ] }_{e_g^*} \\
  \underbrace{[\uparrow, \downarrow] [\uparrow, \downarrow]
  [\uparrow ,\,\,\, ]}_{t_{2g}} 
  \end{array} + \mbox{ligands: } \underbrace{[\,\,\, , \,\,\,] [\,\,\, , \,\,\,] \cdots}_{\pi^*} \, .
  \label{eq:intro.5}
\end{equation}
As this $^3$MC state does not luminesce, its formation quenches the 
phosphoresence of the ruthenium(II) complex leading to shortened luminescence
lifetimes. Here it is important to realize that, 
while we talk of {\em states}, the picture in most
chemical physicists'/physical chemists' minds is that of LFT {\em orbitals}.
However most transition metal complex calculations are carried out using
% -----------------------------
\marginpar{\color{blue} DFT}
% -----------------------------
density-functional theory (DFT) where extracting LFT orbitals is far
from obvious.  Previous work showed how a LFT-like
understanding of ruthenium(II) polypyridine complexes could be 
% -----------------------------
\marginpar{\color{blue} PDOS}
% -----------------------------
obtained using a partial density of state (PDOS) method \cite{WJL+14}, 
in particular allowing assignment of the energy of the antibonding 
$e_g^*$ state which had posed a problem in previous theoretical work.
This, in turn, allowed the construction of an orbital-based luminscence
% -----------------------------
\marginpar{\color{blue} LI3}
% -----------------------------
index of which the third try (LI3) correlated well with 
experimentally-derived $^3$MLCT $\rightarrow$ $^3$MC activation 
energies $E_{\mbox{ave}}$ for about one hundred compounds \cite{MCA+17}.
However we have no reason to believe that $E_{\mbox{ave}}$ is a quantitative
measure of the $^3$MLCT $\rightarrow$ $^3$MC barrier height, even though it
is derived from experiment and so reflects general luminescence trends
in ruthenium(II) polypyridine complexes.  This is further confirmed by
comparison of $E_{\mbox{ave}}$ with the best available experimental and
theoretical values for the same barrier height \cite{MCA+17}.  It thus
makes sense to carry out a more stringent test of LI3 using barriers
calculated at the same level as the original PDOS calculations used to
calculate LI3, namely gas-phase B3LYP/6-31G+LANLDZ(Ru) (see Sec.~\ref{sec:compdetails} for computational details).  Such calculations
are both computer-resource and human-time intensive and so we restricted
our calculations to only a few ruthenium(II) polypyridine complexes.
Several aspects of the results turned out to be rather surprising!
({\em Vide infra!})

\begin{figure}
  \centering
  \includegraphics[width=0.9\textwidth]{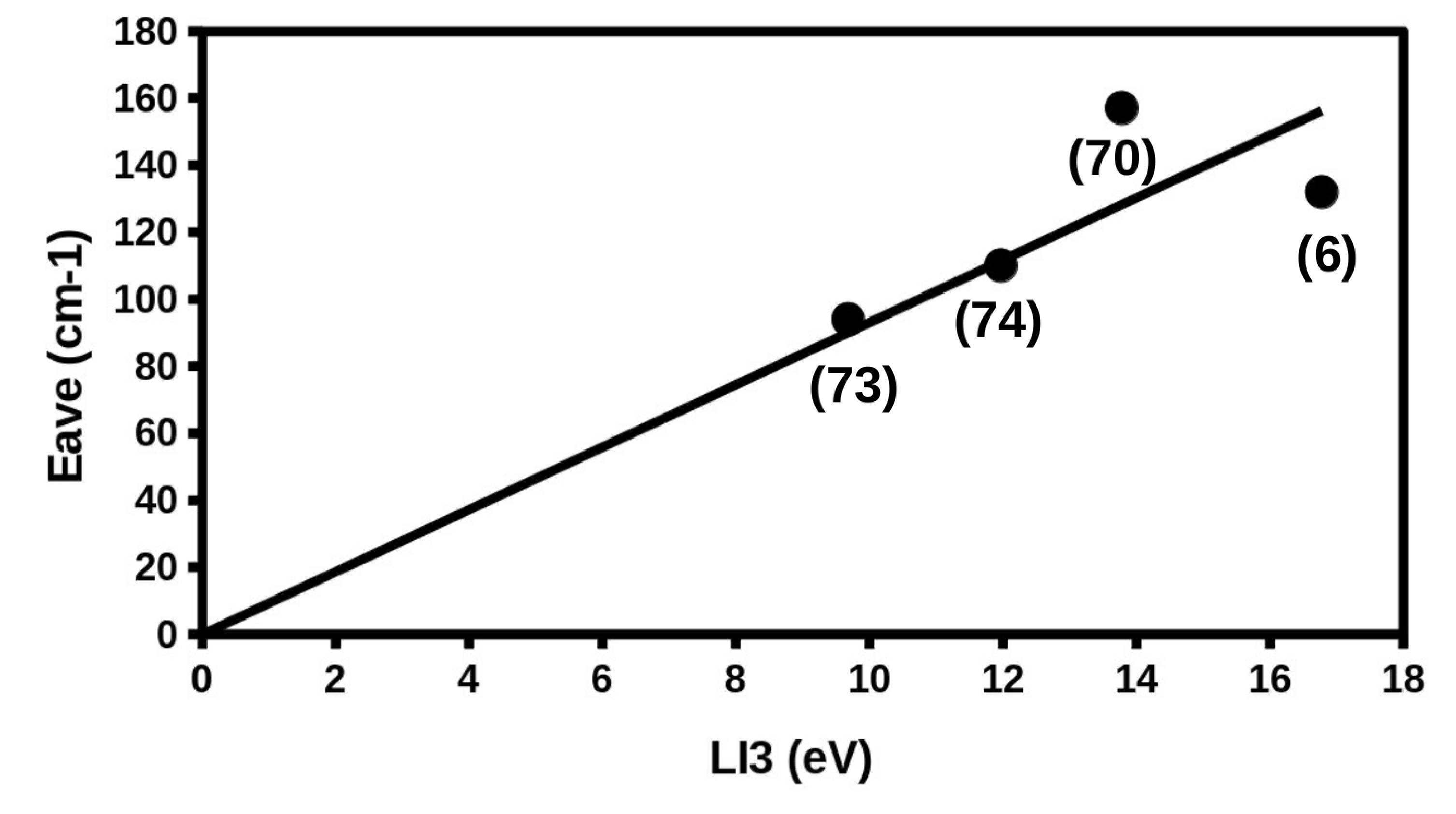}
  \caption{Performance of LI3 for complexes of formula
  [RuX$_3$]$^2+$.  Numbers in parentheses designate the compounds
  shown in {\bf Fig.~\ref{fig:complexstructures}} which, in turn, are named
  after the ligands shown in {\bf Fig.~\ref{fig:ligandlist}}.
  \label{fig:Eave_vs_LI3}
  }
\end{figure}

\begin{figure}
\begin{center}
\begin{tabular}{cc}
  \includegraphics[scale=0.65]{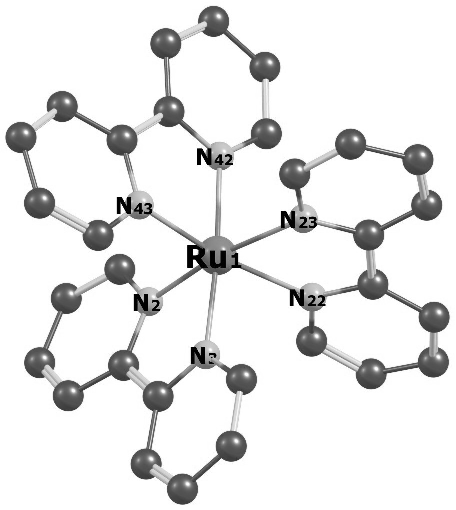}
  &
  \includegraphics[scale=0.65]{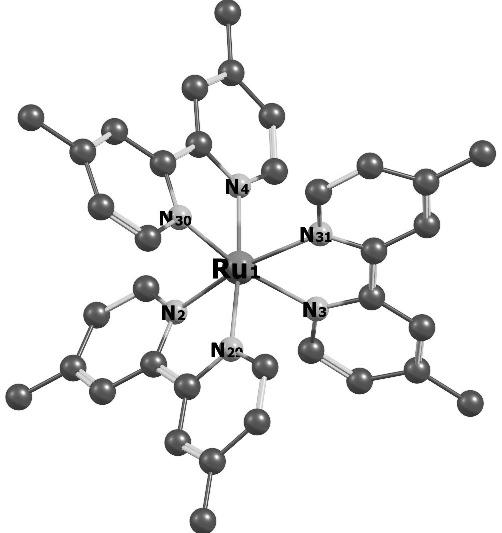}
  \\
  (a) N$^\wedge$N = bpy ({\bf 6})
  &
  (b) N$^\wedge$N = 4,4'-dm-bpy ({\bf 70})
  \\
  \includegraphics[scale=0.65]{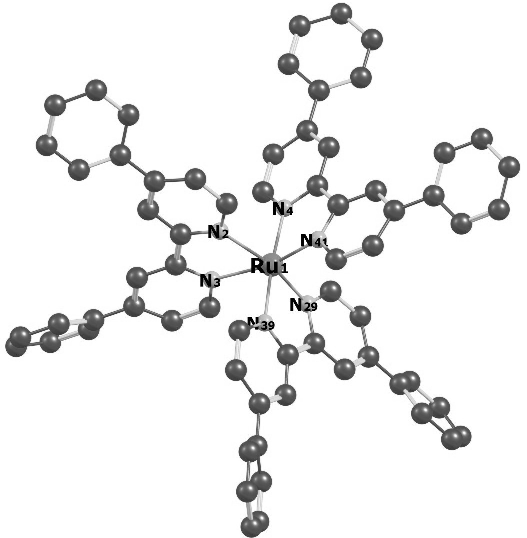}
  &
  \includegraphics[scale=0.65]{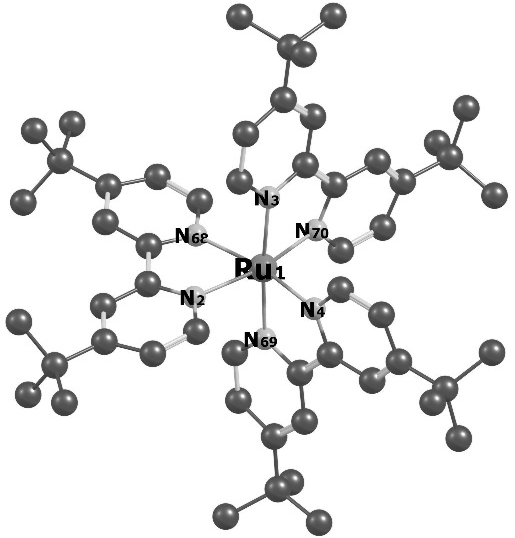}
  \\
  (c) N$^\wedge$N = 4,4'-dph-bpy ({\bf 73})
  &
  (d) N$^\wedge$N = 4,4'-DTB-bpy ({\bf 74})
\end{tabular}
\caption{Structures of [Ru(N$^\wedge$N)$_3$]$^{2+}$.  Hydrogen atoms
have been suppressed for clarity.  We have chosen to use
the $\Delta$ stereoisomers, though similar results are expected
for the corresponding $\Lambda$ stereoisomers.
\label{fig:complexstructures}
}
\end{center}
\end{figure}

\begin{figure}
\begin{center}
\begin{tabular}{c}
  \includegraphics[scale=0.65]{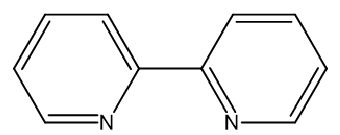}
  \\
  (a) {\bf bpy}: 2,2'-bipyridine ({\bf 6})
  \\
  \includegraphics[scale=0.65]{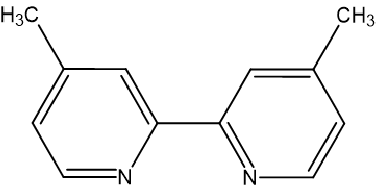}
  \\
  (b) {\bf 4,4'-dm-bpy}: 4,4'-dimethyl-2,2'-bipyridine  ({\bf 70})
  \\
  \includegraphics[scale=0.65]{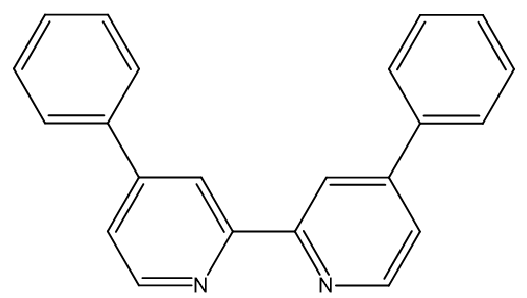}
  \\
  (c) {\bf 4,4'-dph-bpy}: 4,4'-diphenyl-2,2'-bipyridine ({\bf 73})
  \\
  \includegraphics[scale=0.65]{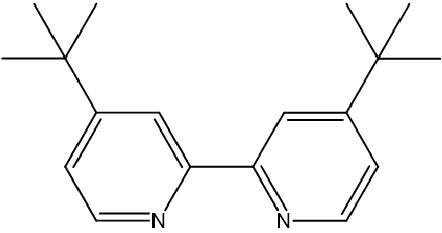}
  \\
  (d) {\bf 4,4'-DTB-bpy}: 4,4'-di-tert-butyl-2,2'-bipyridine ({\bf 74})
\end{tabular}
\caption{Ligand list.
\label{fig:ligandlist}
}
\end{center}
\end{figure}

The $^3$MLCT $\rightarrow$ $^3$MC process may be thought of as
a chemical reaction involving the metal $d$ orbitals and the
ligand $\pi^*$ orbitals.  Just as frontier molecular orbital
% -----------------------------
\marginpar{\color{blue} FMOT}
% -----------------------------
theory (FMOT) has been successful in explaining many chemical reactions,
similar arguments were used to motivate the LI3 orbital-based
luminescence index \cite{MCA+17}
\begin{equation}
  \mbox{LI3}= \frac{\left [ \left ( \epsilon_{{e}_{g}^{*}}
   +\epsilon_{\pi^{*}} \right )/2 \right ]^{2}}
   {\epsilon_{{e}_{g}^{*}}-\epsilon_{\pi^{*}}} \, .
  \label{eq:intro.6}
\end{equation}
Notice how the form of the LI3 descriptor is typical of FMOT with an
orbital energy difference in the denominator.  The numerator represents
the square of the off-diagonal term in the orbital hamiltonian in the 
Helmholtz approximation just as in familiar semi-empirical applications 
of FMOT, but neglecting the orbital overlap term.  As shown in 
{\bf Fig.~\ref{fig:Eave_vs_LI3}}, this descriptor works
remarkably well for predicting $E_{\mbox{ave}}$ for the compounds
considered in the present article.  It still works very well when 
considering a much larger number of compounds though there is necessarily
more scatter in the corresponding graph and a small number of compounds appear 
far enough away from the line that we might start to question whether
they follow the same luminescence mechanism as the other compounds 
\cite{MCA+17}.  This is not entirely surprising because not all bonds are 
% ------------------------------------------------------------------------
% ------------------------------------------------------------------------
symmetry equivalent in many of these compounds. Given the difficulty
of a thorough investigation of reaction barriers, we decided to focus
first on only a small series of similar complexes for which results are reported
in the present article.  These complexes are shown in 
{\bf Fig.~\ref{fig:complexstructures}} and the Lewis dot structures
of the corresponding ligands are shown in {\bf Fig.~\ref{fig:ligandlist}}.
% ----------------------------------------------------------------
% ----------------------------------------------------------------
% -----------------------------------------------------------
This series of complexes was chosen as small and ``simple'' as possible to
% -----------------------------
\marginpar{\color{blue} TS}
% -----------------------------
minimize the problem of multiple transition states (TSs) because locating
TSs and determining their barrier heights can be difficult.

\begin{figure}
  \includegraphics[scale=0.65]{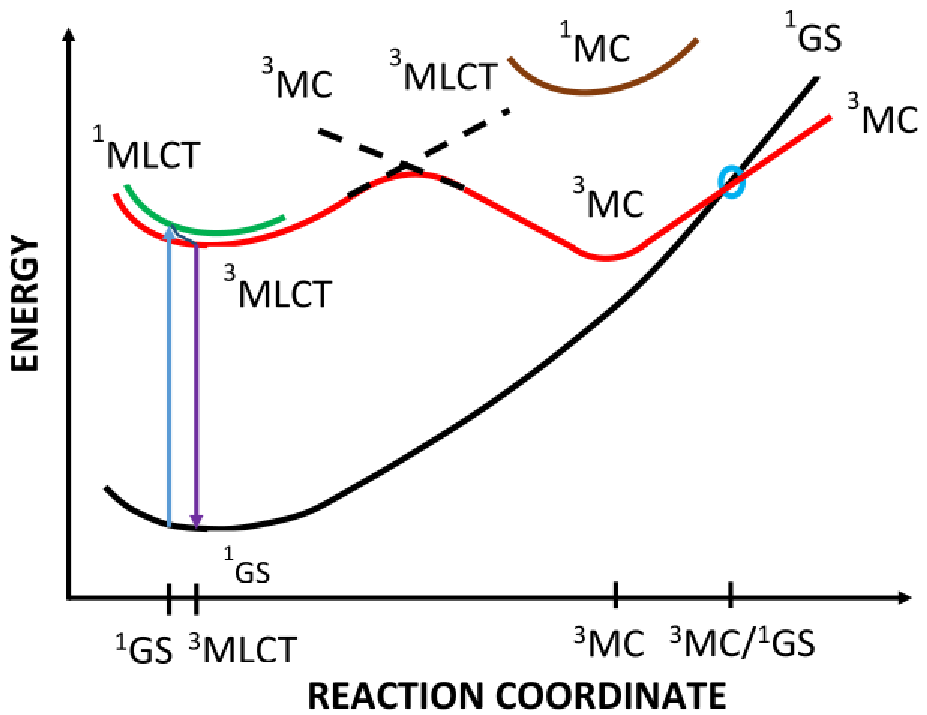}
  \centering
  \caption{The diagram shows the principle potential energy
  curves in our model. The abscissa corresponds to a reaction 
  pathway involving partial removal of a ligand while the ordinate 
  represents the state energy. The dashed lines indicate diabatic 
  states whose avoided crossing leads to the energetic barrier on 
  the adiabatic surface between the $^{3}$MLCT and $^{3}$MC minima. 
  \label{fig:pes}}
\end{figure}

Let us look in more detail at the conventional mechanism for ruthenium(II)
photoluminescence.  On the basis of experimental results and LFT 
group theoretical arguments, Adamson has given a set of rules for 
photodissociation rules for $O_h$ complexes \cite{VC83}:
\begin{quote}
\noindent
{\em Rule 1}. {\em Trans} dissociation will occur for those opposing 
ligands with the weakest average ligand field strength.\\
{\em Rule 2}. In the case of nonidentical ligands, the ligand of greater 
field strength aquates first.
\end{quote}
As Rule 1 appears to be stronger than Rule 2, ruthenium(II) polypyridine
complexes have been almost universally assumed to undergo {\em trans}
dissociation.  This coordinate provides a 1D cut allowing us to sketch
% -----------------------------
\marginpar{\color{blue} PEC}
% -----------------------------
an approximate potential energy curve (PEC).  The luminsence mechanism
described in the first paragraph is then easy to understand in terms
of the diagram in {\bf Fig.~\ref{fig:pes}}.  Note that the $^1$GS
% ---------------------------------------------------------------
% ----------------------------------------------------------------
and $^3$MLCT geometries are represented as very similar, radiationless
$^1$MLCT $\rightarrow$ $^3$MLCT intersystem crossing is very easy because
the two curves have very similar energies and because of spin-orbit
coupling due to the heavy ruthenium atom, the $^1$MC and $^3$MC geometries
are also expected to be similar but with longer ruthenium-ligand bond
lengths because of the occupation of the antibonding $e^*_g$ orbital,
there is an avoided crossing between the $^3$MC and $^3$MLCT states,
and there is easy radiationless de-activation of the $^3$MC state back
to the $^1$GS.

Such a picture is subject to experimental verification but obtaining
barrier heights experimentally requires careful rate measurements at 
a series of temperatures and extraction using an appropriate model, 
usually assuming an Arrhenius rate law.  An accurate experimental
value for the $^3$MLCT $\rightarrow$ $^3$MC barrier height of 
3800 cm$^{-1}$ is known for
the prototypical {\em tris}bipyridine ruthenium(II)
cation complex [Ru(bpy)$_3$]$^{2+}$ in acetonitrile \cite{CM83}
(3960 cm$^{-1}$ in propionitrile/buylronitile (4:5 v/v) is quoted in 
Ref.~\cite{MCA+17}), but for very few other complexes.  The problem 
of obtaining experimental barrier heights has been discussed in more detail
in Ref.~\cite{MCA+17} where a compromise was made in calculating 
% -----------------------------
\marginpar{\color{blue} $E_{\mbox{ave}}$}
% -----------------------------
$E_{\mbox{ave}}$ so as to be able to use data covering a large number
of compounds.  The value $E_{\mbox{ave}}$ = 132 cm$^{-1}$ was found 
for [Ru(bpy)$_3$]$^{2+}$ which clearly underestimates the previously quoted 
experimental barriers, but the hope was that trends in barrier heights
would be accurately captured by the experimentally-derived $E_{\mbox{ave}}$.
This hypothesis is exactly what we wish to test in the present work.

It might seem that determining reaction paths and barrier heights might
be easier than determining them experimentally.  However such complexes
are expected to have a plethera of TSs.  Adamson's {\em trans} dissociation 
of $O_h$ complexes with six identical monodentate ligands is completely
consistent with the expected Jahn-Teller distortion of an open-shell
$O_h$ complex.  However there are three symmetry-equivalent {\em trans} 
dissociation pathways, each of which should be associated with three
% -----------------------------
\marginpar{\color{blue} PES}
% -----------------------------
different TSs on a Mexican hat potential energy (hyper)surface (PES) 
\cite{B01}.  This has also been discussed in the context
of cobalt(II) imine complexes having a single type of bidentate
ligand (e.g., Fig.~7 of Ref.~\cite{KZD+06}).  In addition to the TSs
of each of the three symmetry-equivalent {\em trans} dissociation
products, we may also expect TSs between the symmetry-equivalent
$^3$MLCT geometries and between the symmetry-equivalent $^3$MC
geometries.
In the case of three identical bidentatate ligands, our $O_h$ complex
has $\Lambda$ and $\Delta$ enantiomers.  These also interconvert via
TSs.  Interconversion between the isomers has been proposed to happen
either via Bailar or by Ray-Dutt twists \cite{AJ18}, each of which 
has associated TSs.
And there is the question of whether the excited electron in the 
$^3$MLCT is transferred equally to all three bidentate ligands or
preferentially to a single bidentate ligand?  Certainly photodissociation
experiments which observe the replacement of a bpy ligand with water
in [Ru(bpy)$_3$]$^{2+}$ to form [Ru(bpy)$_2$(H$_2$O)$_2$]$^{2+}$
\cite{VD00} seem consistent with the idea that symmetry breaking is 
occurring in such a way that one of the bpy is treated differently
than the others.  And this reaction also has its own TS.  All of which means
that we must expect many probably close-lying TSs to complicate our
investigation.

Of course we are not the first to carryout a theoretical investigation
of $^{3}$MCLT $\rightarrow$ $^{3}$MC PECs, but the number seems to still
be rather small \cite{YYS+15, ZBP16, ZBP17, SDV+17} and relatively recent.
In their 2015 paper \cite{YYS+15}, Yoshikawa \textit{et al.} report the 
energies and geometries of $^{3}$MCLT $\rightarrow$ $^{3}$MC transition states
for nine [Ru(bpy)$_2$(phen derivative)]$^{2+}$,
where “phen derivative” refers to either (1) phen, (2) 4-phenyl-phen,
(3) 4,7-diphenyl-phen, (4) 2,9-dimethyl-4,7-diphenyl-phen, 
(5) 2,9-dimethyl-phen,(6) 3,4,7,8-tetramethyl-phen, (7) 5-amino-phen,
(8) 5-methyl-phen, and (9) 3-phenylethynyl-phen. However no
PECs were reported. In their 2016 paper \cite{ZBP16, ZBP17},
Zhou \textit{et al.} report $^{3}$MCLT $\rightarrow$ $^{3}$MC PECs
for four derivatives of
fac-tris(1-methyl-5-phenyl-3-n-propyl-[1,2,4]-triazolyl)iridium(III).
In their 2017 paper \cite{SDAH18}, Dixon {\em et al.} report
$^{3}$MCLT $\rightarrow$ $^{3}$MC PECs for
[Ru(bpy)(btz)$_{2}$]$^{2+}$. In their 2017 paper \cite{SDV+17},
Sun \textit{et al.} report $^{3}$MCLT $\rightarrow$ $^{3}$MC
PECs for [Ru(bpy)$_{3}$]$^{2+}$ and for [Ru(mphen)$_{3}$]$^{2+}$.
Finally, in their 2018 papers \cite{SAH+18, SDAH18},
Soupart \textit{et al.} report the $^{3}$MCLT $\rightarrow$ $^{3}$MC
potential energy curves for {\em trans} \cite{SDAH18} and for
{\em cis} \cite{SAH+18} dissociation. Fumanal \textit{et al.}
report transition states for [Mn(im)(CO)$_{3}$(phen)]$^{+}$ but
without detailed potential energy curves \cite{FHG+19}.

Some of the most extensive theoretical work has been done studying
[Ru(bpy)$_3$]$^{2+}$ \cite{AHBV07, HAB09, DHAE17, SDAH18, SAH+18}
where it is clearly seen that the PEC of Fig.~\ref{fig:pes} is an
oversimplification which sould be replaced with a possibly complex
set of competing processes.
To begin with, the $^{3}$MLCT state of [Ru(bpy)$_{3}$]$^{2+}$ has more than 
one minimum, including a minimum with $D_{3}$ symmetry where the ruthenium 
electron has been transferred equally to the three ligands and a minimum 
with $C_{3v}$ symmetry where the ruthenium electron has been transferred 
to a single bpy ligand \cite{AHBV07}. The same reference reports a $^{3}$MC
in order to remove the spatial degeneracy of the $^{3}$MLCT state. 
Spin-orbit effects can explain radiationless relaxation to the ground 
state via S$_{0}/ ^{3}$MC mixing as in shown in Fig.~\ref{fig:pes}
\cite{HAB09}. However LFT considerations \cite{VC83} suggest that 
{\em cis} dissociation might also occur.  Indeed theoretical calculations 
show that {\em cis} triplet dissociation has a comparable but higher barrier 
than does {\em trans} triplet dissociation \cite{DHAE17, SDAH18}. The 
{\em trans} $^{3}$MCLT $\rightarrow$ $^{3}$MC pathway has been mapped out 
in gas phase and solvent for [Ru(bpy)$_{3}$]$^{2+}$ \cite{SAH+18}. 
While the overall conclusion is that complex processes are going on, the
dominant one still seems to be {\em trans} dissociation.
We emphasize that we have consulted with the authors of Ref.~\cite{SAH+18}
and use the same techniques, but have pushed the methodology a little
further in the direction of answering our driving question about LI3
and the energetics of {\em trans} dissociation.

\begin{figure}
  \begin{center}
  % \includegraphics[scale=0.25]{graphics/vri.png}
  % \includegraphics[scale=0.25]{Graphics/vri.eps}
  % \caption{Conceptual diagram of bifurcation and VRI point.
  % Adopted from \cite{vri}. 
  \begin{tabular}{cc}
  (a) & \\
  % (b) & \includegraphics[scale=0.35]{graphics/PESguessNew.eps} \\
  (b) & \includegraphics[scale=0.35]{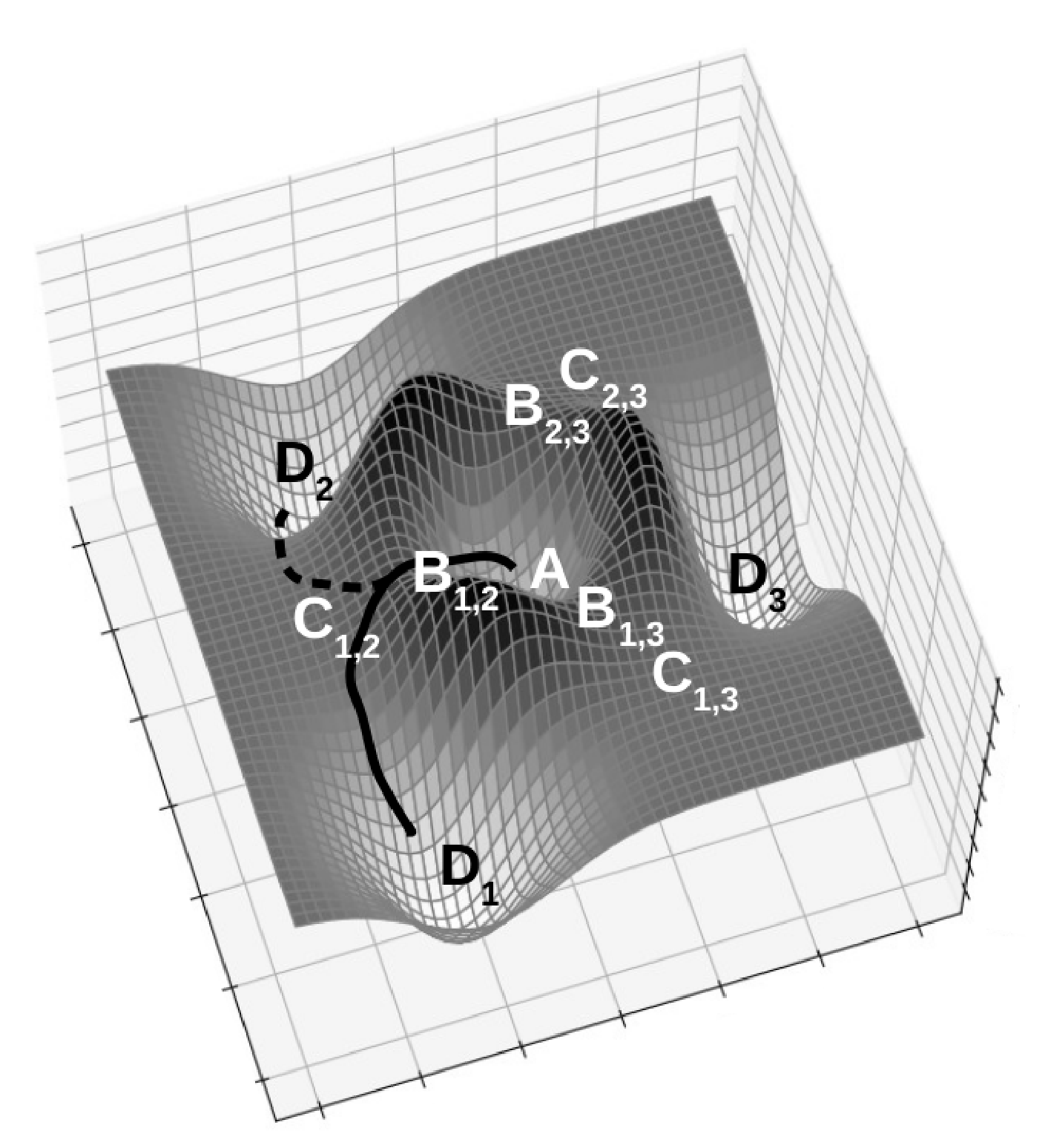} \\
  %    & \includegraphics[scale=0.40]{graphics/PESguessNewB.eps}
      & \includegraphics[scale=0.40]{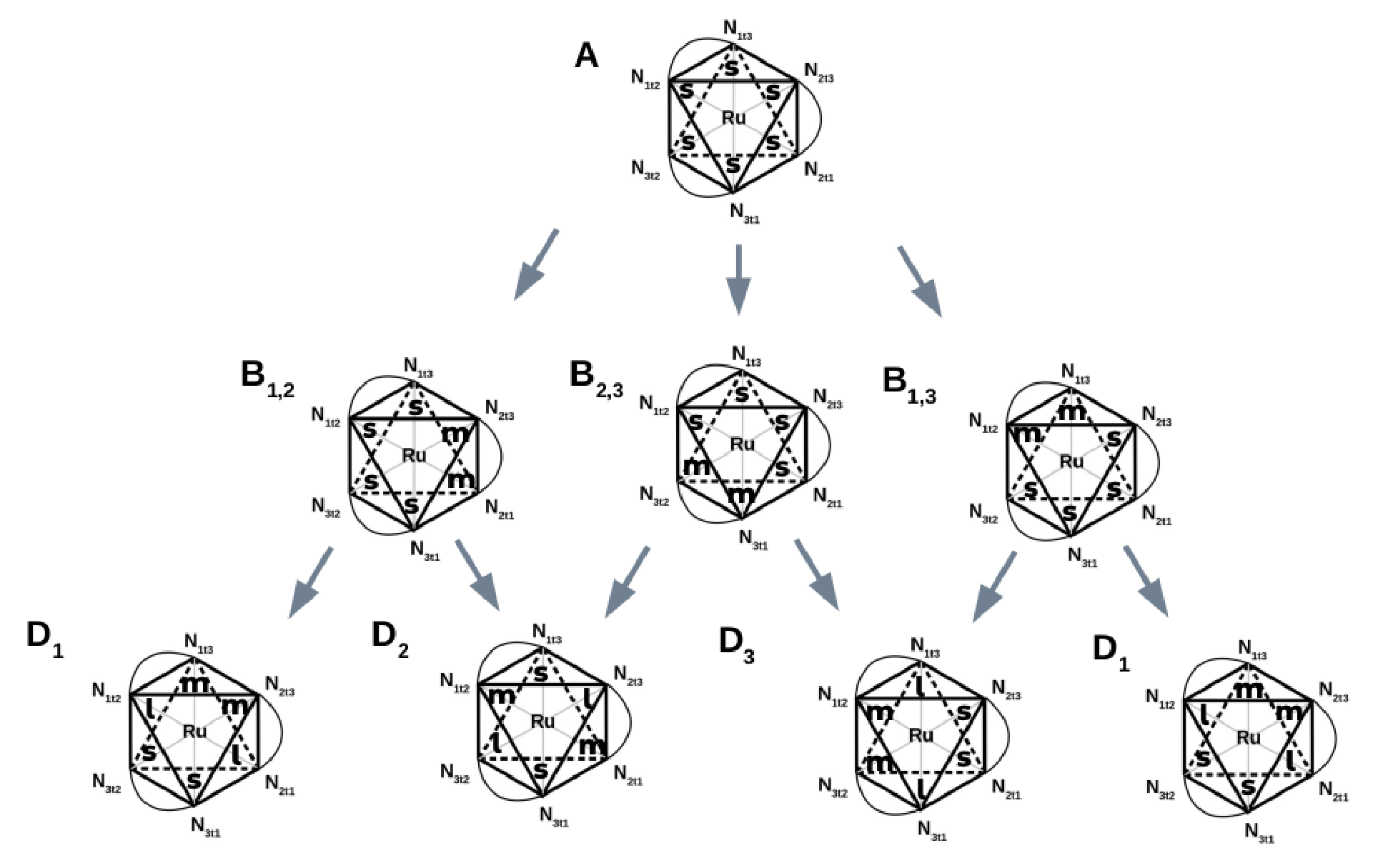}
  \end{tabular}
  \caption{Schematic PES representing our best understanding of the
  $^3$MLCT $\rightarrow$ $^3$MC reaction path: {\bf A} represents three
  nearly identical $^3$MLCT minima, the {\bf B$_{i,j}$} represent three
  TS which lead to a ridge descending to three TSs ({\bf C$_{i,j}$})
  representing the barrier to interconversion between symmetry-equivalent
  $^3$MC minima ({\bf D$_i$} and {\bf D$_j$}). Somewhere along the ridge
  between {\bf B$_{i,j}$} and {\bf C$_{i,j}$} is a valley-ridge inflection
  (VLI) point defined mathematically as the point at which the eigenvalue
  of the Hessian matrix is zero and the gradient vector is perpendicular
  to the corresponding eigenvector) \cite{MHO+15, EWI+08}.  The VLI may
  be thought of as (close to) where the $^3$MLCT $\rightarrow$ $^3$MC
  reaction path bifurcates [part (a) of this figure].
  \label{fig:vri} }
  \end{center}
\end{figure}

Although the present work is restricted to 
a small family of closely related molecules, we have nevertheless realized
that great care was needed in our calculations because something strange
seemed to be happening.  Complex {\bf 6} ([Ru(bpy)$_3$]$^{2+}$) has $N=61$
atoms and therefore $F=3N-6=177$ internal degrees of freedom.  Even though some
of these degrees of freedom are fairly rigid, this still leaves a lot
of possibilities for processes to occur which are different from those initially
imagined!  So we have made a very careful effort to be systematic:  
\begin{enumerate}
  \item Our first step was to make 2D scans of the PES by fixing two 
  {\em trans} bond lengths at different fixed values and minimizing all 
  the other degrees of freedom.  This gives us our first crude view of 
  the reaction mechanism, but scans are well-known to be misleading when 
  it comes to finding TSs and reaction paths (Ref.~\cite{M80} gives a 
  particularly clear explanation of why this is so.)  It does, however,
  provide us with first guesses that are subsequently used to optimize
  the $^3$MC and $^3$MCLT geometries for {\em trans} dissociation.
% -----------------------------
\marginpar{\color{blue} NEB}
% -----------------------------
  \item We then used the nudged elastic band (NEB) method \cite{JMJ98,ABB+21}
  to start from an interpolated path for {\em trans} dissociation and find 
  a first approximation to the reaction path.  This is important as only 
% -----------------------------
\marginpar{\color{blue} MEP}
% -----------------------------
  the NEB, and not the 2D scan, was able to give us a maximum energy point 
  (MEP) close enough to a TS to be useful.
  \item The MEP geometry still has to be optimized using a TS optimizer. 
  Special methods must then be used to locate the TS and
  to determine the reaction pathways \cite{S03, S11}. These are typically
  computationally resource intensive both because they require multiple
  single point quantum chemical calculations but also because of the need
  to calculate first and second derivatives of the PESs
  \cite{S03, S11, PS93, PASF96, HBK05, SZBH12, VR18}. 
  \item The optimized TS still has to be confirmed as linking the $^3$MC
  and $^3$MLCT minima and this is done by following the intrinsic reaction
% -----------------------------
\marginpar{\color{blue} IRC}
% -----------------------------
  coordinate (IRC) in both directions along the direction of the imaginary 
  vibrational coordinate to verify that the TS does indeed correspond to 
  the desired reaction path.
\end{enumerate}
It is also possible to follow the change in character of the molecule along the
IRC by using Mulliken population analysis to follow the spin- and 
charge-density along the reaction path.  Given the surprising nature and
complexity of what we found, we prefer to give a schematic of our main
conclusions and then later justify them.  This schematic PES is shown in 
{\bf Fig.~\ref{fig:vri}}.  The details of the 
notation will be explained later.  Briefly the initial $^3$MLCT state is
quite symmetric so that it is difficult to distinguish three symmetry
equivalent Jahn-Teller distorted geometries.  The IRC is consistent
with an electron being transferred to a particular bidentate ligand whose
two bonds to the central ruthenium atom elongate simultaneously so as
to maintain $C_2$ symmetry at the TS and beyond.  The reaction path is
then on a ridge and can bifurcate in either of two directions to find
one of two symmetry-equivalent minima.  This accounts for small indications
here and there in the literature (and our own experience) that, and 
explains why, the IRC is unusually difficult to calculate for this reaction.
% ----------------------------------------------------------------------
% -----------------------------------------------------------------------

The rest of this paper is organized as follows: Our computational methods
are described in the next section.  Section~\ref{sec:results} justifies
the general picture seen in Fig.~\ref{fig:vri} and then goes on to compare
calculated barrier and reaction energies with LI3.  
Section~\ref{sec:conclusion} summarizes our conclusions.
Additional information is available as Supplementary Information 
% -----------------------------
\marginpar{\color{blue} SI}
% -----------------------------
(SI) (Sec.~\ref{sec:SI}).

% \begin{verbatim}
% 
%  +------+
%  | STOP |
%  +------+
% 
% \end{verbatim}

%%%%%%%
% EOF %
%%%%%%%
% -----------------------------------------------
% \section{Luminescence Indices}
% \label{sec:li3}
% \input{li3.tex}
% -----------------------------------------------
\section{Computational Details}
\label{sec:compdetails}
% \input{compdetails.tex}
% ====================================
% File: compdetails.tex
% Last updated: 30 August 2023
% ====================================

The four-step procedure that was used in the present work has already
been described in the introduction.  This section goes into greater
technical detail.  

We used two programs to carryout
this procedure, namely version 09 revision D.01 of the {\sc Gaussian} 
code \cite{g09} and version 5.0.2 of the {\sc ORCA} code \cite{NWB+20}.
All calculations were gas phase calculations with the same basis sets
and the same functionals.  

% -----------------------------
\marginpar{\color{blue} 6-31G}
% ----------------------------
The basis consisted of the valence double-$\zeta$ quality 6-31G basis
set for H \cite{DHP71}, N \cite{HDP72}, and C \cite{HDP72}. 
The Ru atom requires an effective core
potential (ECP) in order to include relativistic effects.  We used
% -----------------------------
\marginpar{\color{blue} LANL2DZ\\ECP}
% ----------------------------
the LANL2DZ ECP and the corresponding double-$\zeta$ basis set \cite{HW85b}.
% -----------------------------
effective core potential (ECP) \cite{HW85b} was used to characterize Ru.
H, N and C atoms were described by 6-31G basis set. 

% -----------------------------
\marginpar{\color{blue} B3LYP}
% ----------------------------
The density functional approximation used is the B3LYP (Becke exchange, 
3 parameter, Lee-Yang-Parr correlation) functional described
in Ref.~\cite{SDCF94} and initially programmed in {\sc Gaussian}.  
Unfortunately this functional has a history of confusion which can 
% -----------------------------
\marginpar{\color{blue} VWN}
% ----------------------------
be traced back to the implementation of the Vosko-Wilke-Nusair (VWN) 
% -----------------------------
\marginpar{\color{blue} LDA}
% ----------------------------
parameterization of the local density approximation (LDA) in {\sc Gaussian}.
Their parameterization of Ceperley and Alder's quantum Monte Carlo 
results (found in the caption of Fig.~5 of Ref.~\cite{VWN80}) should
have been used as this is what had been referred to as VWN in works
by all previous authors.  However the Vosko-Wilke-Nusair parameterization
% -----------------------------
\marginpar{\color{blue} RPA}
% ----------------------------
of a random phase approximation (RPA) result for the electron gas result
was called VWN in {\sc Gaussian}.  Instead of correcting the error, {\sc
Gaussian} refers to the original parameterization of the Ceperley-Alder
% -----------------------------
\marginpar{\color{blue} VWN5}
% ----------------------------
results as VWN5.  The implementation of B3LYP in {\sc Gaussian} uses
VWN for the LDA.  However B3LYP in {\sc ORCA} uses the orginal VWN (i.e., 
% -----------------------------
\marginpar{\color{blue} B3LYP/G}
% ----------------------------
VWN5) in its implementation of B3LYP and requires the specification of B3LYP/G
in order to get the original B3LYP.  We used the original B3LYP in our 
calculations througout this paper.

While there may be some value in using other functionals and better basis
sets in implicit solvent, such changes would defeat our goal of calculating
barriers to compare with the LI3 indices already calculated at the 
B3LYP/6-31G+LANLDZ(Ru) level used in Ref.~\cite{MCA+17}.  So we decided to
stick with the original level of calculation.

In {\sc Gaussian}, all calculations were carried out without any symmetry
contraints, using an ultrafine grid, and the spin-unrestricted formalism.
% -----------------------------
\marginpar{\color{blue} SCF\\DIIS}
% ----------------------------
Self-consistent field (SCF) convergience was achieved using the direct 
inversion of iterative subspace (DIIS) extrapolation algorithm.
Frequency calculations were 
performed on the optimized geometries so as to check whether they were 
the true minima (with no imaginary frequencies) or a transition state 
(with one imaginary frequency). For all the calculations, ({\tt pop=full}) 
was set to full so as to extract all the information.  
In {\sc ORCA}, keywords used included {\tt NORI} (no approximation is used), 
{\tt TightSCF},  {\tt TightOpt}, {\tt SlowConv}, {\tt NumFreq} and an 
ultra-fine grid. We explicitly confirmed that {\sc Gaussian} and {\sc Orca}
gave the same results for the same basis sets and functionals in single-point
calculations.

% ----------------------------------------------------------------------
\begin{figure}
  \begin{center}
  \includegraphics[scale=0.40]{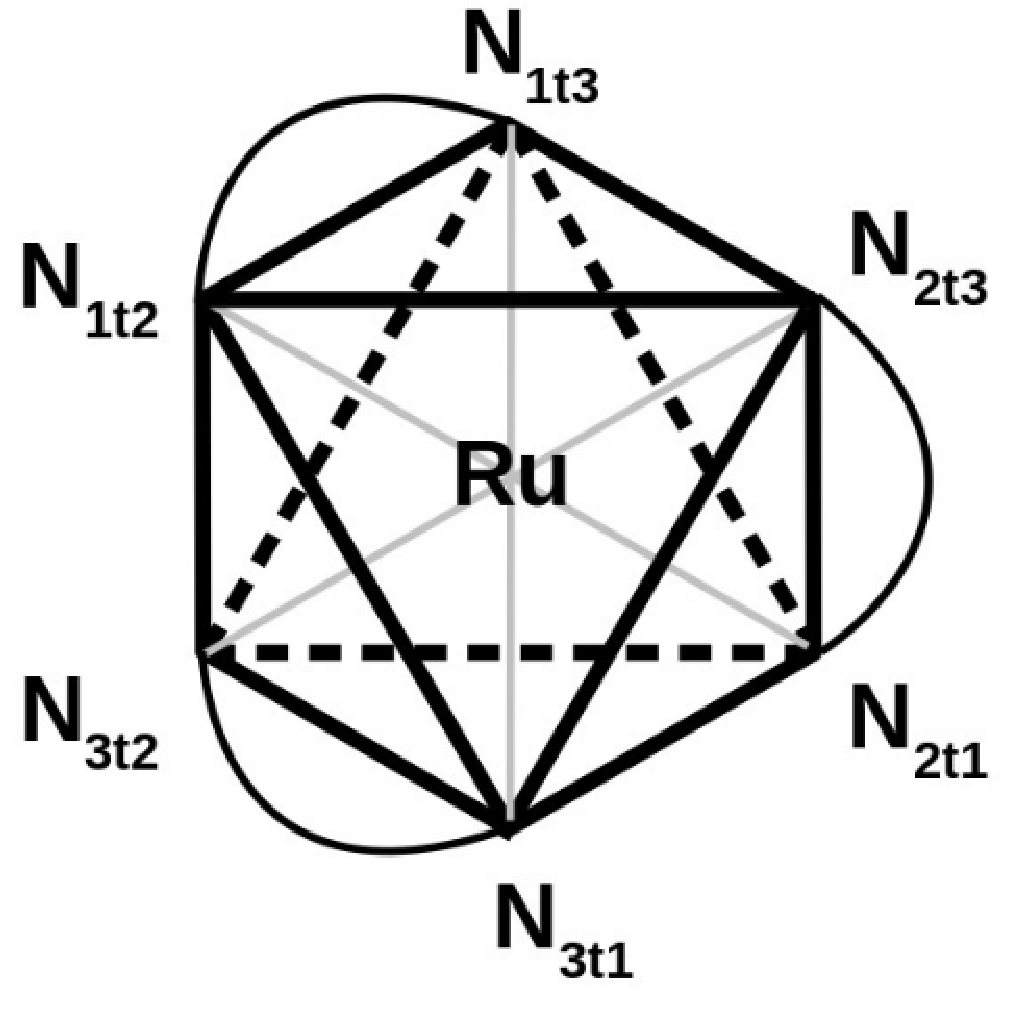}
  \caption{The $m$t$n$ labeling system for the N atoms around the central
  Ru.
  \label{fig:labeling} }
  \end{center}
\end{figure}
% -----------------------------------------------------------------------
We also need some system of N atom numbering in order to be able to 
discuss the photochemical reaction pathways.  One possibility is to
just use the numbers shown in Fig.~\ref{fig:complexstructures}.
However we found that we could be a little less arbitrary by using
a system where each of the N$^\wedge$N ligands
with the numbers 1,2, and 3.  Each of the N in a given N$^\wedge$N is {\em
% -----------------------------
\marginpar{\color{blue} $n$t$m$}
% ----------------------------
trans} to a different N$^\wedge$N, so we may label as $n$t$m$ the $N$ in 
the $n$th N$^\wedge$N that is {\em trans} to the $m$th N$^\wedge$N.  This
is a great improvement over using the atom numbers in 
Fig.~\ref{fig:complexstructures}, but still is complicated by the fact that
there are still six different ways to permute the three numbers (1,2,3).
The possible alternative numberings are shown in {\bf Table~\ref{tab:ntm}}.
With a little practice and an occasional sketch 
({\bf Fig.~\ref{fig:labeling}}), it is relatively easy to 
master.  Usually it is the 2t1-1t2 or the 3t1-1t3 {\em trans} stretch which
interests us in our calculations.  We will now take a step-by-step look at
how we carried out our calculations.  Many of these steps could have been
carried out using either the {\sc Gaussian} or the {\sc ORCA} codes, but 
we chose to allow different people use the codes with which they were most
familiar once we had assured ourselves that the two codes did indeed give the
same results.
% \input{Tables/ntm.tex}
% =======================================
% File: ntm.tex
% Last update: 2 August 2023
% =======================================
% ----------------------------------------------------------------
\begin{table}
\begin{center}
\begin{tabular}{cccccccc}
\hline \hline
Complex & Ru-N Bond  & (1,2,3) & (1,3,2) & (2,1,3) & (2,3,1) & (3,1,2) & (3,2,1) \\
 \hline
[Ru(bpy)$_{3}$]$^{2+}$        & 1-2  & 3t2 & 2t3 & 3t1 & 1t3 & 2t1 & 1t2 \\
      ({\bf 6})             & 1-3  & 3t1 & 2t1 & 3t2 & 1t2 & 2t3 & 1t3 \\
                            & 1-22 & 2t1 & 3t1 & 1t2 & 3t2 & 1t3 & 2t3 \\
                            & 1-23 & 2t3 & 3t2 & 1t3 & 3t1 & 1t2 & 2t1 \\
                            & 1-42 & 1t3 & 1t2 & 2t3 & 2t1 & 3t2 & 3t1 \\
                            & 1-43 & 1t2 & 1t3 & 2t1 & 2t3 & 3t1 & 3t2 \\
\mbox{[}Ru(4,4'-dm-bpy)$_3$]$^{2+}$  & 1-2  & 3t2 & 2t3 & 3t1 & 1t3 & 2t1 & 1t2 \\
     ({\bf 70})      	    & 1-3  & 2t1 & 3t1 & 1t2 & 3t2 & 1t3 & 2t3 \\
                     	    & 1-4  & 1t3 & 1t2 & 2t3 & 2t1 & 3t1 & 3t1 \\
                     	    & 1-29 & 3t1 & 2t1 & 3t2 & 3t2 & 2t3 & 1t3 \\
                     	    & 1-30 & 1t2 & 1t3 & 2t1 & 2t3 & 2t1 & 3t2 \\
                     	    & 1-31 & 2t3 & 1t3 & 1t3 & 3t1 & 1t2 & 2t1 \\
\mbox{[}Ru(4,4'-dph-bpy)$_3$]$^{2+}$ & 1-2  & 1t2 & 1t3 & 2t1 & 2t3 & 3t1 & 3t2 \\
     ({\bf 73})      	    & 1-3  & 1t3 & 1t2 & 2t3 & 2t1 & 3t2 & 3t1 \\
                     	    & 1-4  & 3t2 & 2t3 & 3t1 & 1t3 & 2t1 & 1t2 \\
                     	    & 1-29 & 2t1 & 3t1 & 1t2 & 3t2 & 1t3 & 2t3 \\
                     	    & 1-39 & 2t3 & 3t2 & 1t3 & 3t1 & 2t1 & 2t1 \\
                     	    & 1-41 & 3t1 & 2t1 & 3t2 & 1t2 & 2t3 & 1t3 \\
\mbox{[}Ru(4,4'-DTB-bpy)$_3$]$^{2+}$ & 1-2  & 3t1 & 2t1 & 3t2 & 1t2 & 2t3 & 1t3 \\
     ({\bf 74})      	    & 1-3  & 1t2 & 1t3 & 2t1 & 2t3 & 3t1 & 3t2 \\
                     	    & 1-4  & 2t3 & 3t2 & 1t3 & 3t1 & 2t1 & 2t1 \\
                     	    & 1-68 & 3t2 & 2t3 & 3t1 & 1t3 & 2t1 & 1t2 \\
                     	    & 1-69 & 2t1 & 3t1 & 1t2 & 3t2 & 1t3 & 2t3 \\
                     	    & 1-70 & 1t3 & 1t2 & 2t3 & 2t1 & 3t2 & 3t1 \\
\hline \hline
\end{tabular}
\caption{The six different ways to relabel N atoms as in the $n$th
N$^\wedge$N unit {\em trans} to the $m$th N$^\wedge$N unit.
With a little practice, this notation becomes fairly easy to  use.
\label{tab:ntm}
}
        \end{center}
\end{table}
% ---------------------------------------------------------------

\paragraph{Step 1} consists of making a 2D scan of the PES.  This was done
using {\sc Gaussian} with the objective of finding a suitable $^3$MLCT minimum
energy geometry.  This was achieved by first doing a vertical excitation
% -----------------------------
\marginpar{\color{blue} FC}
% ----------------------------
from the $^1$GS minimum to obtain the Franck-Condon (FC) $^3$MLCT.  This
FC $^3$MLCT geometry was then optimized to get the $^3$MLCT local minimum.
(It is {\em not} a global minimum!) This minimum was confirmed 
by checking for the absence of imaginary frequencies.   We then made
a simultaneous 2D-scan by independently stretching two {\em trans} Ru-N
bonds and made a contour plot.  This strategy was adopted from \cite{AHBV07} 
which suggested that the ground state molecular orbitals are of the 
$d_{z^2}$-type and the population of this $d_{z^2}$-like orbital in 
the triplet excited state would thus result in the bond elongations (i.e., 
the $e^*_g$ LFT orbital is antibonding).  Of course all three {\em trans} 
distortions should give the same reaction path.
Two axial bonds that were {\em trans} to each other 
were independently elongated from the initial bond length in the 
optimized geometry of the $^{3}$MLCT state to 
2.500 {\AA} in steps of 0.002 {\AA}, letting all other geometric parameters
relax to give the lowest energy with only the two {\em trans} Ru-N distances
constrained.  The resulting energies had to be resorted before plotting because
{\sc Gaussian} uses a boustrophedon (as an ox plows a field) order when it 
does a 2D-scan in order to minimize changes in geometries during the scan.
The lowest energy point with long {\em trans} bonds is a first guess at
the $^3$MC geometry.  This was then optimized once again to obtain the 
minimum energy geometry and the minima was confirmed by checking for
the absence of any imaginary frequencies. Mulliken spin density analysis 
was used to confirm of the nature of the excited-state \{0.9 excess spin-up
electron on Ru [Eq.~(\ref{eq:intro.2})] for the $^{3}$MLCT state and 
1.8 excess spin-up electron 
on Ru for a $^{3}$MC state [Eq.~(\ref{eq:intro.3})]\}.     

\paragraph{Step 2} Having found our $^3$MLCT and $^3$MC end points, we 
then carried out an NEB calculation \cite{JMJ98, ABB+21} with {\sc ORCA} 
to find a best first guess at the IRC.  

\paragraph{Step 3} The MEP from the NEB calculation in {\sc ORCA} was further
optimized using {\sc ORCA}'s NEB-TS algorithm.  The resultant TS was reoptimized
with a different algorithm with {\sc Gaussian}.

\paragraph{Step 4} Finally {\sc Gaussian} was used to follow the IRC from the
calculated TS and insure that it is indeed connected to the originally input
$^3$MLCT and $^3$MC input geometries.

% ================================================
\section{Results and Discussion}
\label{sec:results}
% \input{results.tex}
% =======================================
% File: results.tex
% Last update: 6 September 2023
% The energy gap (3MLCT → 3MC state) is an important parameter that 
% determines the probability of ic (3MLCT → 3MC state). If the 3MC 
% state of a Ru-polypyridine complex is lower in energy than its 
% 3MLCT state, ic (3MLCT → 3MC state) can take place easily, being 
% favorable for PCT. On the other hand, if a complex has a higher 
% 3MC state over the 3MLCT state, the probability of ic (3MLCT → 3MC state) 
% would decrease and would have more potential for PDT. 
% https://doi.org/10.1021/acsomega.3c01006
% =======================================

We are now prepared to investigate the question initially posed, namely how 
well the orbital-based luminescence index LI3 predicts the 
$^3$MLCT $\rightarrow$ $^3$MC barrier height.  
A first subsection will focus on the mechanism and energetics of the 
$^3$MLCT $\rightarrow$ $^3$MC reaction.  The method of calculation of
the barrier height was explained earlier as was the reason for the choice
of the level of our calculations to be consistent with our earlier calculations
of LI3.  While we will focus on results for complex {\bf 70} (the results 
for the other three complexes are in the SI), we also want to reveal the
correctness of the schematic PES of Fig.~\ref{fig:vri} as we believe this
to be a new and novel contribution to the literature of this well-studied
and important family of molecules.  A second, and final, subsection will then
examine the question of what we can say about the predictive value of LI3.

% -------------------------------------------------------------------
\subsection{$^3$MLCT $\rightarrow$ $^3$MC Mechanism and Energetics}
% -------------------------------------------------------------------

In the first instance, we focus on complex {\bf 70} and show the results
for the various steps of our reaction mechanism investigation.

% --------------------------------------------------------
\begin{figure}
  \centering
  \includegraphics[scale=0.45]{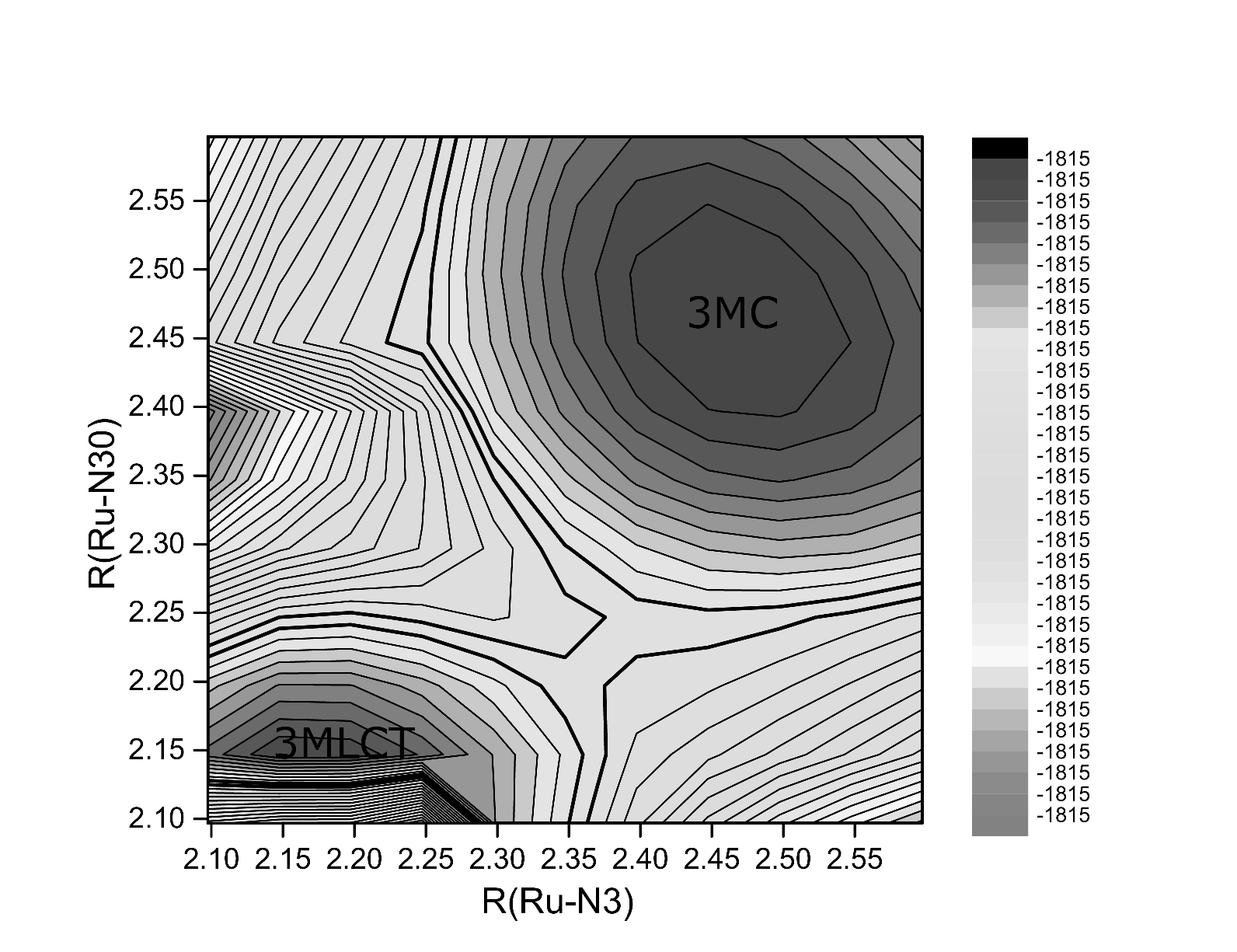}
  \caption{Contour plot of 2D-Scan of Ru$_{1}$-N$_{3}$ (2t1) and 
   Ru$_{1}$-N$_{30}$ (1t2) for complex {\bf 70}. 
  \label{fig:my_label_70}
  }
\end{figure}  
% --------------------------------------------------------
\paragraph{Step 1} The calculated contour plot is shown in 
{\bf Fig.~\ref{fig:my_label_70}}.  Inspection of a model shows
that this plot should be symmetric about the 45$^\circ$ line representing
interchange between the 2t1 and 1t2 N atoms.  It is approximately but not
exactly symmetric because of the existance of multiple internal degrees
of freedom in this system which may be trapped in different local minima 
due to the boustrophedon nature of {\sc Gaussian}'s 2D scan.  This phenomenon
is well understood \cite{M80}.  Nevertheless the contour plot is nearly
symmetric.  It allows us to see that there is an energetically high-lying
$^3$MLCT minimum and an energetically lower-lying $^3$MC minimum, separated
by a barrier.  Importantly, this allows us to find a first guess of the
$^3$MC structure which is subsequently optimized.  Key optimized Ru-N
bond lengths are collected in {\bf Table~\ref{tab:optimized_geometriesA}}.
% \input{Tables/optimized_geometriesA.tex}
% =======================================
% File: optimized_geometriesA.tex
% Last updated: 6 September 2023
% =======================================
\begin{table}
\begin{center}
\begin{tabular}{ccccccccc}
\hline \hline
Ru-N & \multicolumn{2}{c}{Ground State} 
      & \multicolumn{2}{c}{$^3$MLCT}                  
        & \multicolumn{2}{c}{TS}                             & \multicolumn{2}{c}{$^3$MC} \\
Bond    & {\sc Gaussian} & {\sc Orca}  
        & {\sc Gaussian} & {\sc Orca}
        & {\sc Gaussian} & {\sc Orca}
        & {\sc Gaussian} & {\sc Orca} \\
\hline
\multicolumn{9}{c}{[Ru(bpy)$_3$]$^{2+}$ ({\bf 6})}\\
% Ru$_1$-N$_2$ 
3t2          & 2.110 (s) & 2.111 (s)
             & 2.096 (s) & 2.098 (s)
             & 2.101 (s) & 2.100 (s)
             & 2.115 (s) & 2.115 (s) \\
% Ru$_1$-N$_3$ 
3t1          & 2.110 (s) & 2.111 (s)
             & 2.097 (s) & 2.098 (s)
             & 2.102 (s) & 2.102 (s)
             & 2.112 (s) & 2.115 (s) \\
% Ru$_1$-N$_{22}$ 
2t1          & 2.110 (s) & 2.110 (s)
             & 2.118 (s) & 2.119 (s)
             & 2.211 (m) & 2.214 (m)
             & 2.463 (l) & 2.464 (l) \\
% Ru$_1$-N$_{23}$ 
2t3          & 2.110 (s) & 2.110 (s)
             & 2.118 (s) & 2.119 (s)
             & 2.215 (m) & 2.212 (m)
             & 2.176 (m) & 2.170 (m) \\ 
% Ru$_1$-N$_{42}$ 
1t3          & 2.110 (s) & 2.110 (s)
             & 2.097 (s) & 2.099 (s)
             & 2.103 (s) & 2.102 (s)
             & 2.164 (m) & 2.170 (m) \\
% Ru$_1$-N$_{43}$ 
1t2          & 2.110 (s) & 2.110 (s)
             & 2.095 (s) & 2.097 (s)
             & 2.098 (s) & 2.101 (s)
             & 2.454 (l) & 2.464 (l) \\
\multicolumn{9}{c}{[Ru(4,4'-dm-bpy)$_3$]$^{2+}$ ({\bf 70})}\\
% Ru$_1$-N$_2$ 
3t2          & 2.109 (s) & 2.109 (s)
             & 2.097 (s) & 2.097 (s)
             & 2.098 (s) & 2.096 (s)
             & 2.129 (s) & 2.112 (s) \\
% Ru$_1$-N$_3$ 
2t1          & 2.109 (s) & 2.109 (s)
             & 2.097 (s) & 2.097 (s)
             & 2.227 (m) & 2.225 (m)
             & 2.437 (l) & 2.465 (l) \\
% Ru$_1$-N$_{4}$ 
1t3          & 2.109 (s) & 2.109 (s)
             & 2.097 (s) & 2.097 (s)
             & 2.103 (s) & 2.103 (s)
             & 2.139 (m) & 2.186 (m) \\
% Ru$_1$-N$_{29}$ 
3t1          & 2.109 (s) & 2.109 (s)
             & 2.097 (s) & 2.097 (s)
             & 2.103 (s) & 2.104 (s)
             & 2.104 (s) & 2.112 (s) \\ 
% Ru$_1$-N$_{30}$ 
1t2          & 2.109 (s) & 2.109 (s)
             & 2.097 (s) & 2.097 (s)
             & 2.098 (s) & 2.097 (s) 
             & 2.387 (l) & 2.465 (l) \\
% Ru$_1$-N$_{31}$ 
2t3          & 2.109 (s) & 2.109 (s)
             & 2.097 (s) & 2.097 (s)
             & 2.226 (m) & 2.225 (m)
             & 2.222 (m) & 2.168 (m) \\
\multicolumn{9}{c}{[Ru(4,4'-dph-bpy)$_3$]$^{2+}$ ({\bf 73})}\\
% Ru$_1$-N$_2$ 
1t2          & 2.104 (s) & 2.104
             & 2.114 (s) & 2.094 (s)
             & 2.251 (m) & 2.251 (m)
             & 2.165 (m) & 2.459 (l)     \\
% Ru$_1$-N$_3$ 
1t3          & 2.104 (s) & 2.104
             & 2.115 (s) & 2.091 (s)
             & 2.258 (m) & 2.258 (m)
             & 2.458 (l) & 2.165 (m)      \\
% Ru$_1$-N$_{4}$ 
3t2          & 2.104 (s) & 2.104
             & 2.091 (s) & 2.098 (s)
             & 2.103 (s) & 2.103 (s)
             & 2.165 (m) & 2.108 (s)      \\
% Ru$_1$-N$_{29}$ 
2t1          & 2.104 (s) & 2.104
             & 2.087 (s) & 2.094 (s)
             & 2.099 (s) & 2.099 (s)
             & 2.108 (s) & 2.457 (l) \\ 
% Ru$_1$-N$_{39}$ 
2t3          & 2.104 (s) & 2.104
             & 2.094 (s) & 2.091 (s)
             & 2.106 (s) & 2.106 (s)
             & 2.108 (s) & 2.165 (m) \\
% Ru$_1$-N$_{41}$ 
3t1          & 2.104 (s) & 2.104
             & 2.086 (s) & 2.098 (s)
             & 2.101 (s) & 2.101 (s)
             & 2.456 (l) & 2.108 (s) \\
\multicolumn{9}{c}{[Ru(4,4'-DTB-bpy)$_3$]$^{2+}$ ({\bf 74})}\\
% Ru$_1$-N$_2$ 
3t1          & 2.107 (s) & 2.107
             & 2.130 (s) & 2.130 (s)
             & 2.229 (m) & 2.230 (m)
             & 2.357 (l) & 2.465 (l) \\
% Ru$_1$-N$_3$ 
1t2          & 2.107 (s) & 2.107
             & 2.092 (s) & 2.092 (s)
             & 2.102 (s) & 2.102 (s)
             & 2.106 (s) & 2.166 (s) \\
% Ru$_1$-N$_{4}$ 
2t3          & 2.107 (s) & 2.107
             & 2.075 (s) & 2.073 (s)
             & 2.090 (s) & 2.090 (s)
             & 2.210 (m) & 2.110 (m) \\
% Ru$_1$-N$_{68}$ 
3t2          & 2.107 (s) & 2.107
             & 2.130 (s) & 2.130 (s)
             & 2.231 (m) & 2.231 (m)
             & 2.357 (l) & 2.167 (m) \\ 
% Ru$_1$-N$_{69}$ 
2t1          & 2.107 (s) & 2.107
             & 2.092 (s) & 2.092 (s)
             & 2.102 (s) & 2.102 (s)
             & 2.105 (s) & 2.109 (s)        \\
% Ru$_1$-N$_{70}$ 
1t3          & 2.107 (s) & 2.107
             & 2.075 (s) & 2.073 (s)
             & 2.090 (s) & 2.090 (s)
             & 2.210 (m) & 2.459 (l) \\
\hline \hline
\end{tabular}
\caption{
Key Ru-N bond lengths ({\AA}) for different compounds as computed with 
{\sc Gaussian} and with {\sc ORCA}. In parentheses: ``s'' stands for ``short''
($\sim$ 2.0 {\AA}),
``m'' for ``medium length'' ($\sim$ 2.1 {\AA}), and ``l'' for ``long'' 
($\sim$ 2.4 {\AA}).
\label{tab:optimized_geometriesA}
}
        \end{center}
\end{table}

% ---------------------------------------------------------------------
\begin{figure}
  \begin{center}
  \includegraphics[scale=0.95]{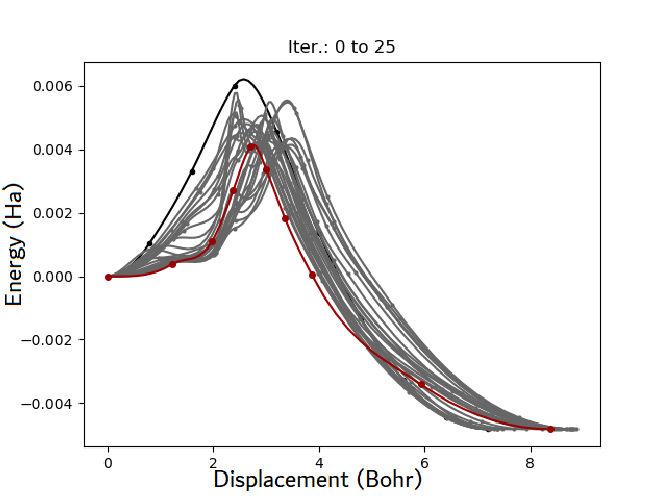}
  \caption{NEB-TS optimization for complex {\bf 70}. 
  The initial guess IRC is in black, the final converged IRC is in red, 
  and the intermediate IRCs obtained during the NEB procedure are shown
  in grey.
  \label{fig:nebopt_70} }
  \end{center}
\end{figure}  
% --------------------------------------------------------------------
\paragraph{Step 2} {\bf Figure~\ref{fig:nebopt_70}} shows the iterative
convergence of the NEB IRC from its initial guess to its final result. 
This is an expensive calculation involving the simultaneous displacement
of 24 points (the end points are fixed) in a parallelized calculation.
It is clear that its convergence is approximate.

% ---------------------------------------------------------------------------
\begin{figure}
  \centering
  \includegraphics[scale=0.45]{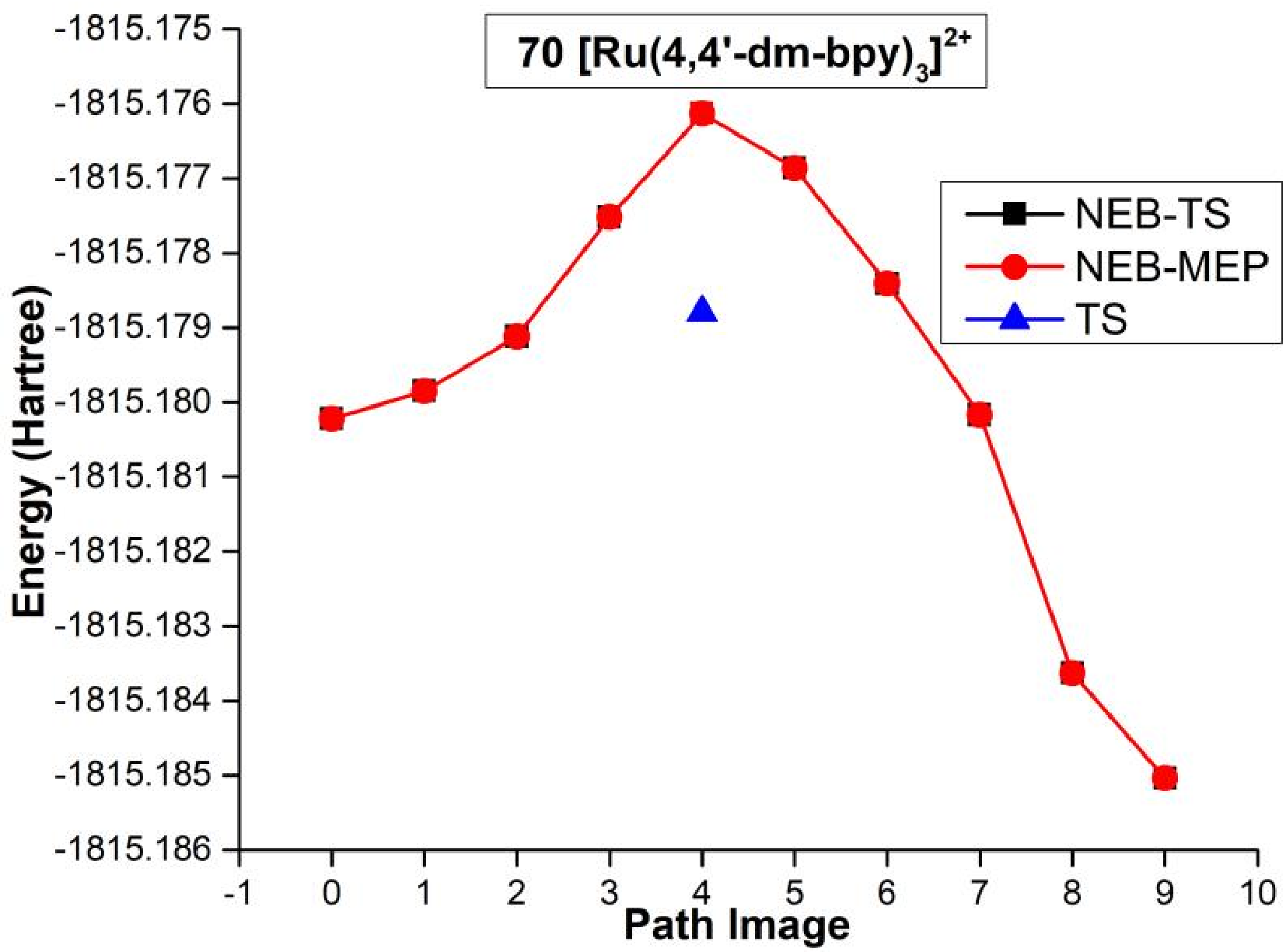}
  \caption{Plot of the NEB IRC for complex {\bf 70}. The red circles are on top
  of the black squares.  The blue point is the energy of the optimized TS.
  \label{fig:70_neb_irc_vs_energy_MEP}
  }
\end{figure}
% -------------------------------------------------------------------------
\paragraph{Step 3} {\bf Figure~\ref{fig:70_neb_irc_vs_energy_MEP}} shows the
IRC obtained from the NEB calculation.  In this particular case, the MEP
optimizes to a point which gives a considerably lower barrier.  The 
geometries of the optimized TSs obtained with {\sc Gaussian} and {\sc Orca}
may compared in Table~\ref{tab:optimized_geometriesA}. Agreement is to within
0.002 {\AA}.

% --------------------------------------------------------------
\begin{figure}
  \centering
  \includegraphics[width=0.9\textwidth]{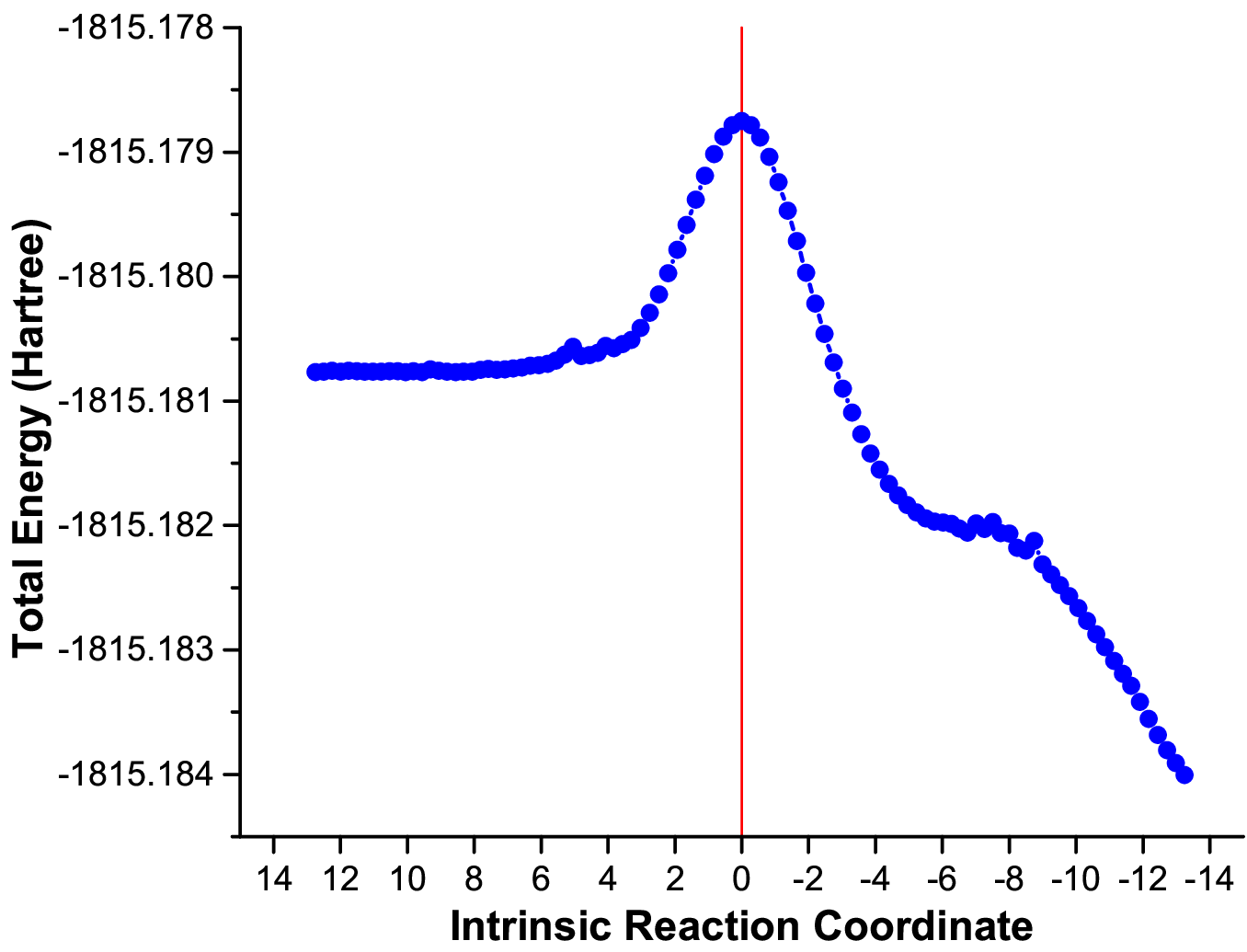}
  \caption{Plot of the {\sc Gaussian} total energy in Hartree against the
  IRC for complex {\bf 70}. 
  \label{fig:IRC_against_energy_compound_70}
  }
\end{figure}
% --------------------------------------------------------------
\paragraph{Step 4} {\bf Figure~\ref{fig:IRC_against_energy_compound_70}}
shows the more accurate IRC obtained using {\sc Gaussian}.  Note that
this figure evolves from left to right in the $^3$MLCT $\rightarrow$ $^3$MC
direction while Fig~\ref{fig:70_neb_irc_vs_energy_MEP} evolves from left
to right in the $^3$MC $\rightarrow$ $^3$MLCT direction.  These calculations
confirmed that the TS is connected by the IRC to the originally input 
$^3$MC and $^3$MLCT structures.

% --------------------------------------------------------------
\begin{figure}
  \centering
  \includegraphics[scale=0.45]{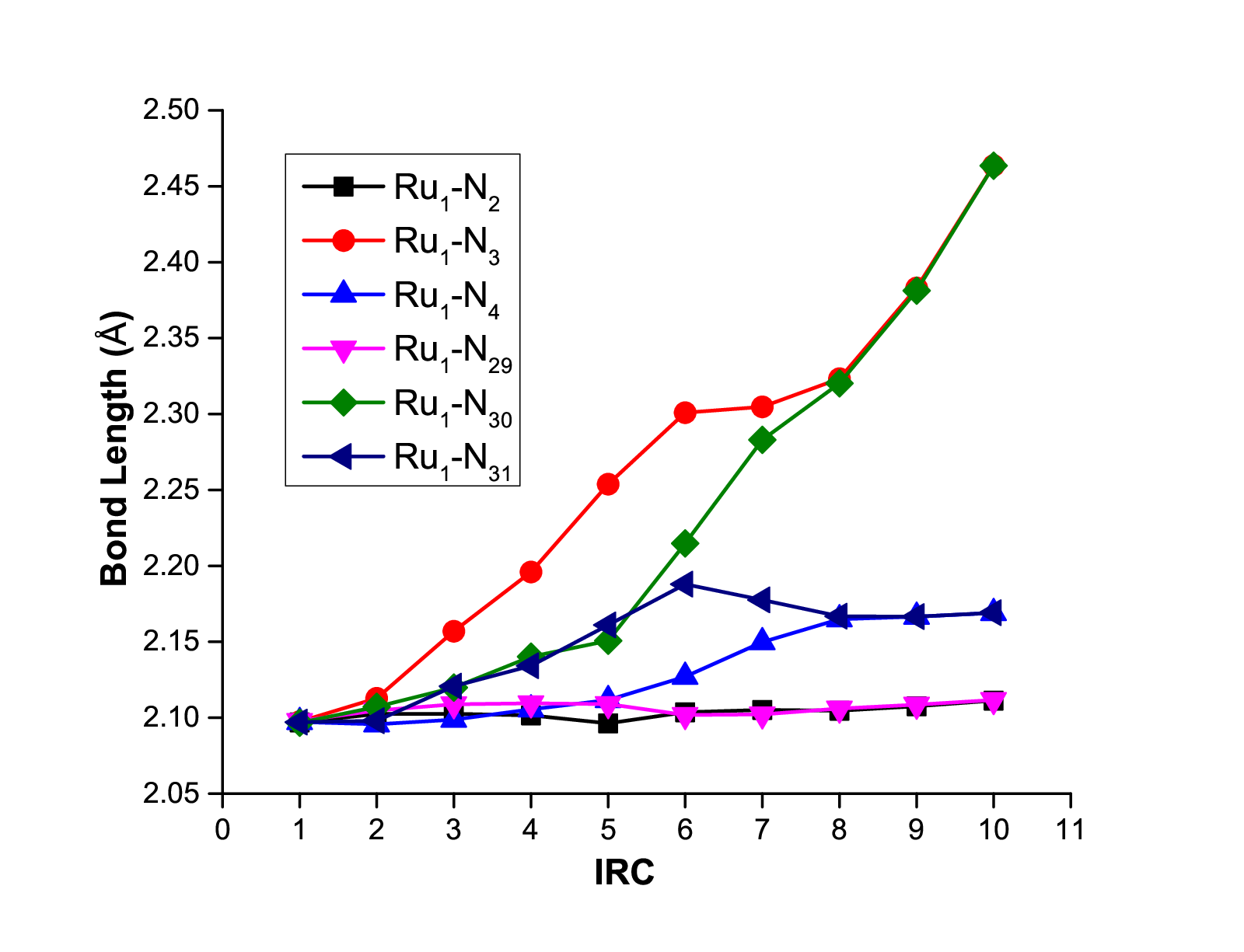}
  \caption{Plot of bond lengths along the IRC for complex {\bf 70} as 
  calculated with {\sc Orca}.
  \label{fig:70_neb_irc_vs_bond_length}
  }
\end{figure}
% --------------------------------------------------------------
\begin{figure}
  \centering
  \includegraphics[scale=0.45]{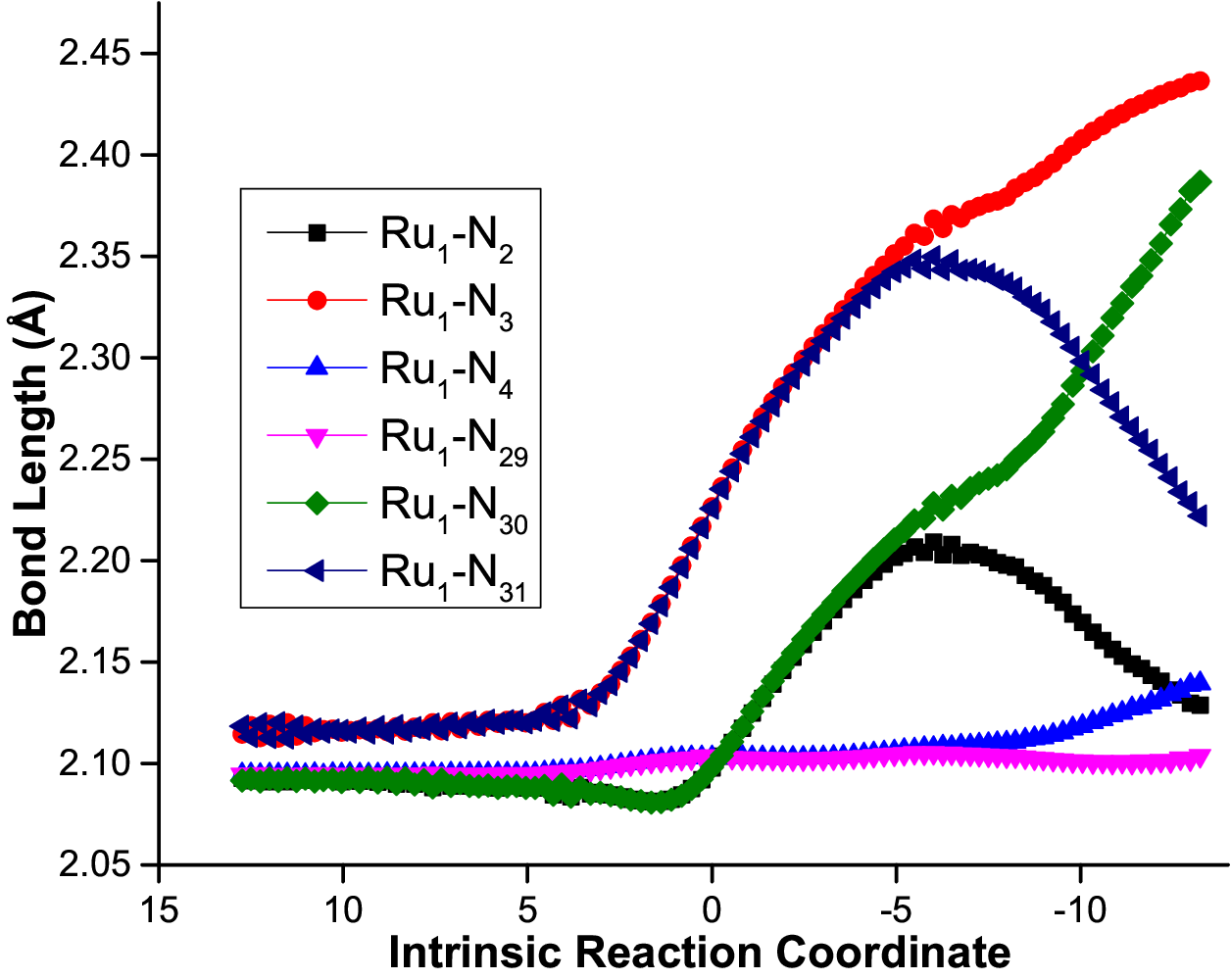}
  \caption{Plot of bond lengths along the IRC for complex {\bf 70} as 
  calculated with {\sc Gaussian}.
  \label{fig:Bond Length_compound_70}
  }
\end{figure}
% ------------------------------------------------------------------
\paragraph{Reaction Mechanism} We have also looked at spin densities, but we
have found the evolution of Ru-N bond lengths along the IRC to be the
most valuable indicator of how the reaction mechanism proceeds.
{\bf Figures~\ref{fig:70_neb_irc_vs_bond_length}} and 
{\bf \ref{fig:Bond Length_compound_70}} show how these bond lengths
vary along the IRC during the photoreaction.  The NEB calculation 
(Fig.~\ref{fig:70_neb_irc_vs_bond_length}) from left to right follows
the $^3$MC $\rightarrow$ $^3$MLCT reaction.  It is less accurate than
the IRC calculation carried out with {\sc Gaussian} 
(Fig.~\ref{fig:Bond Length_compound_70}) which from left to right follows
the $^3$MLCT $\rightarrow$ $^3$MC direction.  Only the {\sc Gaussian} IRC
is accurate enough to understand what is going on.  From right to left: 
The initial $^3$MLCT Ru-N bond lengths are nearly equal until the IRC 
is reduced to about 3, at which point the 2t1 and 2t3 bond lengths remain
equal but increase while the other bond lengths continue to stay about the 
same.  This means that ligand 2 is moving away from the metal in a way that 
might be expected if the $e_g^*$  electron on Ru had been transferred to a 
$\pi^*$ orbital on ligand 2.  This continues beyond the TS (at IRC
coordinate 0 in Fig.~\ref{fig:Bond Length_compound_70}) until we get to 
an IRC coordinate of about -5.  So, up to this point, one of the ligands
is coming off by elongating both of its Ru-N bonds equally.  When the 
IRC is less than -5, there is a sudden change in the bond lengths, so that
the two {\em trans} Ru-N pairs (2t1 and 1t2) become longest while two other
pairs (3t2 and 2t2) reduce to a medium bond length.  Two Ru-N pairs
(3t2 and 3t1) remain short.  Such behavior is consistent with the
PES scheme shown in Fig.~\ref{fig:vri}.  Such a PES should show a bifurcation
such that different {\em trans} bond lengths could lengthen in different
calculations.  This is seen in Table~\ref{tab:optimized_geometriesA} for 
complexes {\bf 73} and {\bf 74}, confirming our overall interpretation 
of the {\em trans} dissociation reaction mechanism.

% -------------------------------------------------------------------
\subsection{What Does LI3 Predict?}
% -------------------------------------------------------------------

We established in our earlier work that the orbital-based luminescence
index LI3 correlates well with $E_{\mbox{ave}}$ values of the 
$^3$MLCT $\rightarrow$ $^3$MC barrier derived from experimental luminescence
data obtained at room temperature and at the boiling point of liquid nitrogen
\cite{MCA+17}.  This has been illustrated in Fig.~\ref{fig:Eave_vs_LI3}.
{\em The question is not {\em whether} LI3 works (it does!) but
how it is able to do this.} We will address this question more fully in 
this subsection.

% =======================================
% File: lifetimes.tex
% Last update: 4 August 2023
% =======================================
% ----------------------------------------------------------------
\begin{table}
\begin{center}
\begin{tabular}{cccc}
\hline \hline
Complex & $\tau$/$\mu$s (77K) & $\tau$/$\mu$s (RT) 
& $\Delta E_{\mbox{ave}}$/cm$^{-1}$ \\
 \hline
({\bf 6})  & 5.23 & 0.845 & 132. \\
({\bf 70}) & 4.6  & 0.525 & 157. \\ 
({\bf 73}) & 4.79 & 1.31  & 94. \\
({\bf 74}) & 9.93 & 0.673 & 194. \\
\hline \hline
\end{tabular}
\caption{Luminescence lifetimes at both room temperature (RT) and
liquid nitrogen temperature (77K) along with the empirically-derived
$\Delta E_{\mbox{ave}}$ as reported in Tables 10 and 11 of 
Ref.~\cite{MCA+17}.
\label{tab:lifetimes}
}
        \end{center}
\end{table}
% ---------------------------------------------------------------

%%%%%%%
% EOF %
%%%%%%%
It is perhaps appropriate to recall that the fundamental 
$^3$MLCT $\rightarrow$ $^3$MC reaction involves an electron transfer.
One famous theory of electron transfer is Marcus theory \cite{M56} and
Marcus theory has indeed been used in connection with studying
electrogenerated chemiluminescence in tris(polypyridine) ruthenium
complexes \cite{KB99}.  Marcus theory is valid when the Massey parameter,
\begin{equation}
  \Gamma = \frac{2\pi \vert H_{1,2} \vert^2 }{\vert d(E_1-E_2)/dt \vert} \, ,
  \label{eq:results.1}
\end{equation}
is small, where $E_1$ and $E_2$ are the energies of the two diabatic curves
($^3$MLCT and $^3$MC here) and $H_{1,2}$ is the adiabatic coupling.  
Diagonalizing the corresponding $2 \times 2$ matrix gives the adiabatic curves,
corresponding to the PESs treated in the present work.  Marcus theory assumes
rapid crossing of the avoided crossing region in the weak coupling case (i.e.,
when $E_1 \approx E_2$).  A detailed derivation \cite{Bertsch} shows that 
Marcus theory assumes rapid oscillation between the reactants and 
products.  The final formula for the reaction rate constant, expressed 
in terms of Gibb's free energies $G$ is,
\begin{equation}
  k = \vert G_{R,P} \vert^2 \sqrt{\frac{\pi}{\lambda k_B T}} 
  e^{-\frac{\left( \lambda + \Delta G^0 \right)^2}{4\lambda k_B T}} \, ,
  \label{eq:results.2}
\end{equation}
where the relaxation parameter $\lambda$ is the difference between the
product diabatic curve at the reactant minimum and the same curve at the
product minimum.  In contrast, the strong coupling limit combined
with a no-recrossing rule, leads to Eyring's transition state theory 
\cite{LK83} and the equally well-known formula,
\begin{equation}
  k = \frac{\kappa k_B T}{2\pi} e^{-{\frac{\Delta G^\ddagger}{RT}}} \, ,
  \label{eq:results.3}
\end{equation}
which explains the famous Arrhenius equation,
\begin{equation}
  k = A e^{-\frac{E_a}{RT}} \, ,
  \label{eq:results.4}
\end{equation}
familiar from first-year University chemistry courses.  Marcus theory
and Eyring transition state theory are, in fact, just two limiting cases
of the more general theory of charge transfer reactions \cite{BMR96}.
We have not defined every parameter in these well-known equations because
we do not need them in the present work and because we are sure that the 
interested reader can easily find suitable references to fill in any 
missing details.  But
it is evident that temperature $T$ is in these equations which may at first
seem a little strange for a photochemical reaction.  A justification is
that the luminescence lifetimes of our complexes are on the order of
microseconds ({\bf Table~\ref{tab:lifetimes}}) whereas a typical vibrational 
time is on the order of picoseconds, which suggests that some degree of 
thermodynamic equilibration may occur, even if it is not necessarily
complete.  In turn, this explains why the usual analysis of 
barriers for these complexes is carried out in terms of the Arrhenius equation 
with some additional terms involving equilibrium between near-lying electronic 
states and a melting term (see Ref.~\cite{MCA+17} and references therein).  
In our case, {\em we are only interested in the energetics of the 
$^3$MLCT $\rightarrow$ $^3$MC reaction and the extent to which it correlates 
with LI3}.  We will assume the Arrhenius picture with the same freqency
factor $A$ for all our complexes so that only the activation energy
$E_a$ is important.  Of course, we are also limited by the fact that we
are using the same level of calculation of the PESs that we used to
calculate our LI3s.  We have to accept in advance that this level of 
calculation may or may not be adequate for directly explaining the 
experimental observations, even if we expect it to shed light upon our
problem.

% \input{Tables/Table4.tex}
% =======================================
% File: Table4.tex
% Last updated: 31 August 2023
% =======================================
    
\begin{center}
\begin{table}
\footnotesize
\begin{tabular}{ c c c c c }
\hline \hline
Energy{\textbackslash}Complex    & {\bf 6}           & {\bf 70}          & {\bf 73}          & 
                       {\bf 74}          \\
\hline
LI3                  & 16.78 eV          & 13.78 eV          & 9.68 eV           & 
                       11.97 eV          \\
$^{3}$MLCT           & -1579.29913 Ha    & -1815.18022 Ha    & -2965.37532 Ha    & 
                       -2522.64515 Ha    \\
$^{3}$MC             & -1579.30474 Ha    & -1815.18504 Ha    & -2965.37816 Ha    &
                       -2522.64934 Ha    \\
$^{3}$MEP            & -1579.29661 Ha    & -1815.17613 Ha    & -2965.37120 Ha    &
                       -2522.64201 Ha    \\
$^{3}$TS             & -1579.29769 Ha    & -1815.17880 Ha    & -2965.37392 Ha    &
                       -2522.64328 Ha    \\
$^{3}$MEP-$^{3}$MLCT & 0.00252 Ha        & 0.00409 Ha        & 0.00412 Ha        &
                       0.00314 Ha        \\
                     & 553.08 cm$^{-1}$  & 897.65 cm$^{-1}$  & 904.24 cm$^{-1}$  &
                       689.15 cm$^{-1}$  \\
$^{3}$TS-$^{3}$MLCT  & 0.00144 Ha        & 0.00142 Ha        & 0.0014 Ha         &
                       0.00187 Ha        \\
                     & 316.04 cm$^{-1}$  & 311.65 cm$^{-1}$  & 307.26 cm$^{-1}$  &
                       410.42 cm$^{-1}$  \\
$^{3}$MEP-$^{3}$MC   & 0.00813 Ha        & 0.00891 Ha        & 0.00696 Ha        &
                       0.00733 Ha        \\
                     & 1784.33 cm$^{-1}$ & 1955.52 cm$^{-1}$ & 1527.54 cm$^{-1}$ &
                       1608.75 cm$^{-1}$ \\
$^{3}$TS-$^{3}$MC    & 0.00705 Ha        & 0.00624 Ha        & 0.00424 Ha        &
                       0.00606 Ha        \\
                     & 1547.52 cm$^{-1}$ & 1369.52 cm$^{-1}$ & 930.57 cm$^{-1}$ &
                       1330.02 cm$^{-1}$ \\
$^{3}$MC-$^{3}$MLCT  & 0.005610 Ha       & 0.00482 Ha        & 0.00284 Ha        &
                       0.004190 Ha       \\
                     & 1231.25 cm$^{-1}$ & 1057.87 cm$^{-1}$ & 623.31 cm$^{-1}$  &
                       919.60 cm$^{-1}$  \\
\hline \hline 
 \end{tabular}
 \caption{Energies for $^{3}$MLCT, $^{3}$MC and TS of 4 compounds 
 obtained from NEB calculations as implemented in the \textsc{ORCA} code. 
        \label{tab:orcaenergies}}
\end{table}
\end{center}

%%%%%%%
% EOF %
%%%%%%%
{\bf Table~\ref{tab:orcaenergies}} gives energies for the $^{3}$MLCT 
states, $^{3}$MC states, and TSs for the four compounds obtained from NEB 
calculations as implemented in the \textsc{Orca} code. Values in 
parenthesis are in cm$^{-1}$.  Except for the $^3$MEP values which 
come from NEB calculations, the same numbers are obtained
with {\sc Gaussian}.  It is immediately clear that the
$^3$MLCT $\rightarrow$ $^3$MC barrier ($^3$MEP-$^3$MLCT or, for
more accuracy, $^3$TS-$^3$MC) is really very small and on the order
of about 300-400 cm$^{-1}$ (0.86-1.14 kcal/mol).  Chemical accuracy is
usually cited as 1 kcal/mol (349.757 cm$^{-1}$) and is very hard to 
achieve even with the best quantum chemical methods.  

% --------------------------------------------------------------
\begin{figure}
  \centering
  \includegraphics[scale=0.45]{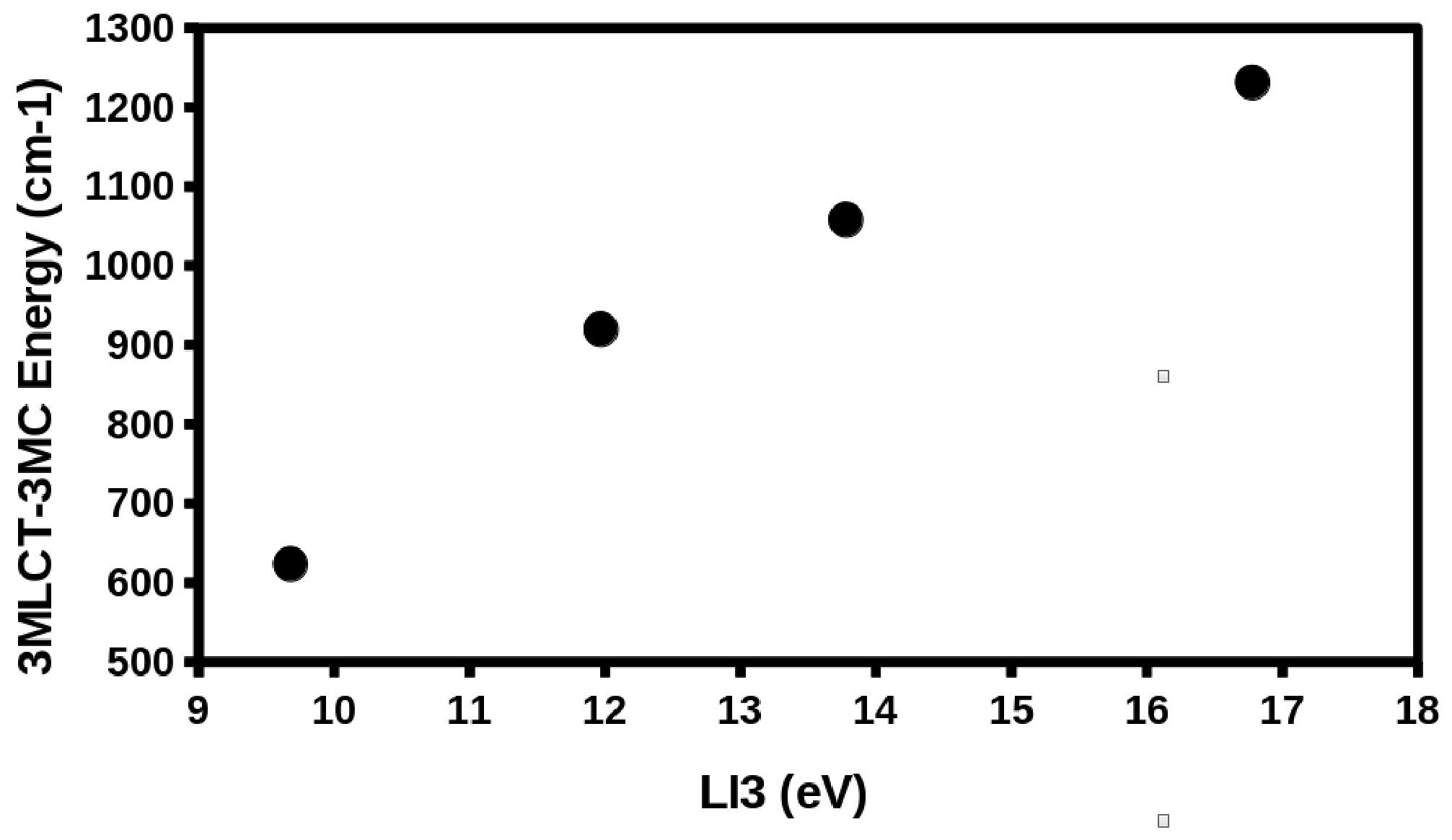}
  \caption{Graph of the $^3$MLCT-$^3$MC energy difference 
  versus LI3.
  \label{fig:DeltaEvsLI3}
  }
\end{figure}
% ------------------------------------------------------------------
{\bf Figure~\ref{fig:DeltaEvsLI3}} shows that the $^3$MLCT-$^3$MC 
energy difference correlates very well with L3---in fact even 
better than the correlation of $\Delta E_{\mbox{ave}}$ with LI3
(Fig.~\ref{fig:Eave_vs_LI3}). This is consistent with the origin of
the FMOT origin of the orbital-based luminescence index LI3.

% --------------------------------------------------------------
\begin{figure}
  \centering
  \includegraphics[scale=0.45]{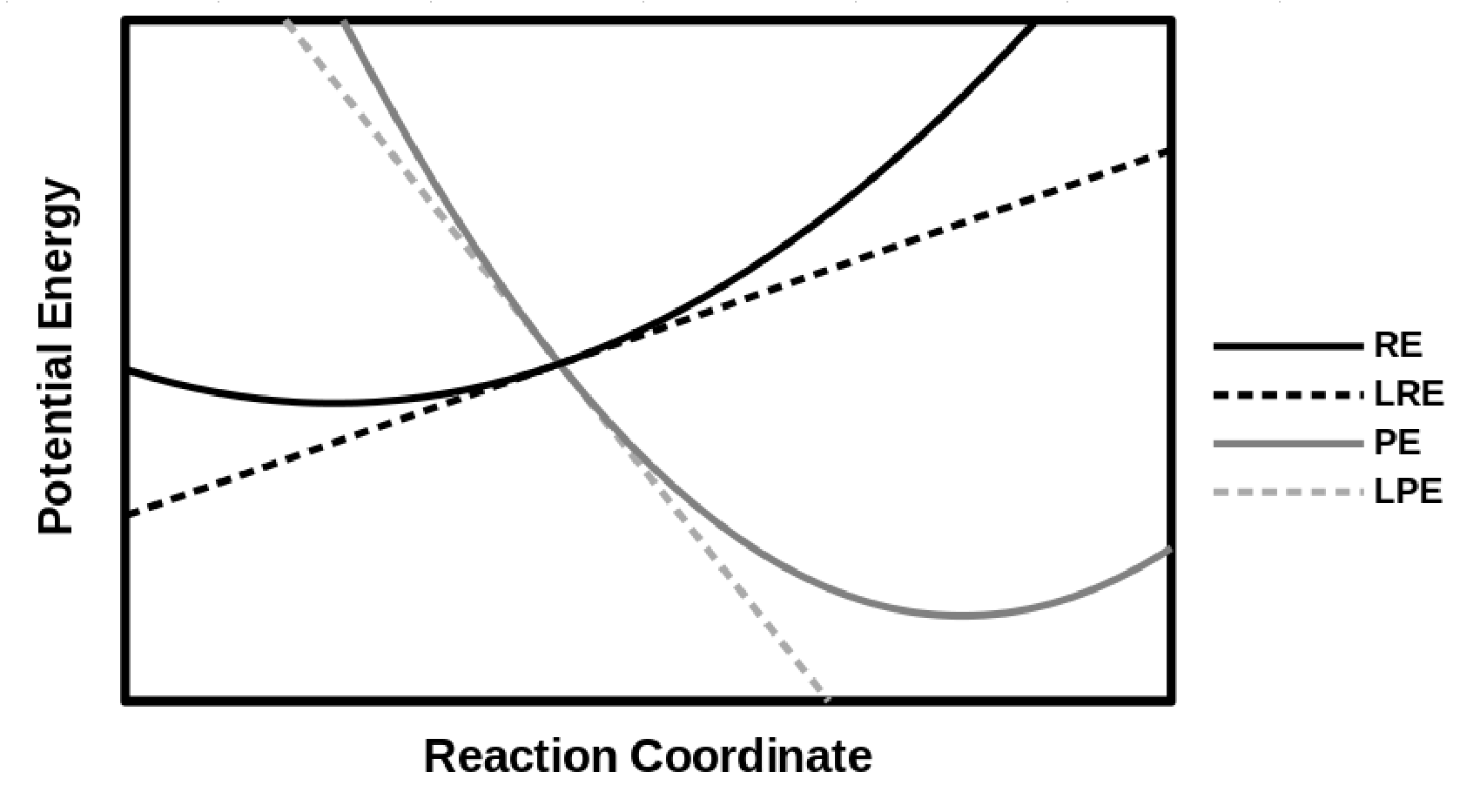}
  \caption{Explanation of the Bell-Evans-Polanyi model \cite{B36,EP36}
  in the present context: RE, reactant PEC, $E_R(x) = (k_R/2)(x-x_R)^2+E_R^0$;
  LRE, linearized RE, $E_R(x) = m_R (x-x_R) + E_R^0$; PE, product PEC,
  $E_P(x)= (k_P/2)(x-x_P)^2+E_P^0$; LPE, linearized PE, $E_P(x) = m_P (x-x_P)
  + E_P^0$.  Here $x$ is the reaction coordinate.
  versus 
  \label{fig:BEP}
  }
\end{figure}
% ------------------------------------------------------------------
There is an argument that the $^3$TS-$^3$MLCT energy barrier should vary 
linearly with the $^3$MLCT-$^3$MC energy difference, 
\begin{equation}
  E[\mbox{$^3$TS}] - E[\mbox{$^3$MLCT}] =
  \alpha \left( E[\mbox{$^3$MC}] - E[\mbox{$^3$MLCT}] \right) + E_0 \, ,
  \label{eq:results.5}
\end{equation}
and hence that the $^3$TS-$^3$MLCT energy barrier should also correlate 
quite nicely with L3.  This can be understood in terms of two parabollic
intersecting diabatic curves as shown in {\bf Fig.~\ref{fig:BEP}}.  The
intersection point $(x_a,E_a+E_R^0)$ defines the activation energy ($E_a$).
The corresponding reaction coordinate may be found by solving the equation,
\begin{equation}
  x_a = \frac{x_P + x_R}{2} + \frac{\Delta E^0 + \left(\frac{\Delta k}{2}\right)
  (x_a-x_R)^2}{k_P(x_P-x_R)} \, ,
  \label{eq:results.6}
\end{equation}
where $\Delta E^0 = E_P^0-E_R^0$ and $\Delta k = k_P-k_R$.  
Equation~(\ref{eq:results.6}) may be solved for $x_a$ in many different ways,
but taking $\Delta k=0$ as an initial guess and then interating leads to 
rapid convergence in the case we tried. Back in 1936, Bell, Evans, and Polanyi
\cite{B36,EP36} presented an argument which is easily used in the present 
context.  It requires linearlization of the parabolas (dotted lines in 
Fig.~\ref{fig:BEP}) and then solving, which gives,
\begin{equation}
  E_a = \underbrace{\frac{m_R}{m_R-m_P}}_{\alpha} \Delta E^0
  + \underbrace{E_R^0 + \frac{m_R m_P}{m_P-m_R} (x_P - x_R)}_{E_0} \, .
  \label{eq:results.7}
\end{equation}
Identification of the reactant as the $^3$MLCT state and the product as the
$^3$MC state, then gives Eq.~(\ref{eq:results.5}).  Solving with the full
parabolas (which is a step in the derivation of Marcus theory) would lead
to an additional quadratic term, which we will ignore here.

\begin{figure}
  \centering
  \includegraphics[scale=0.45]{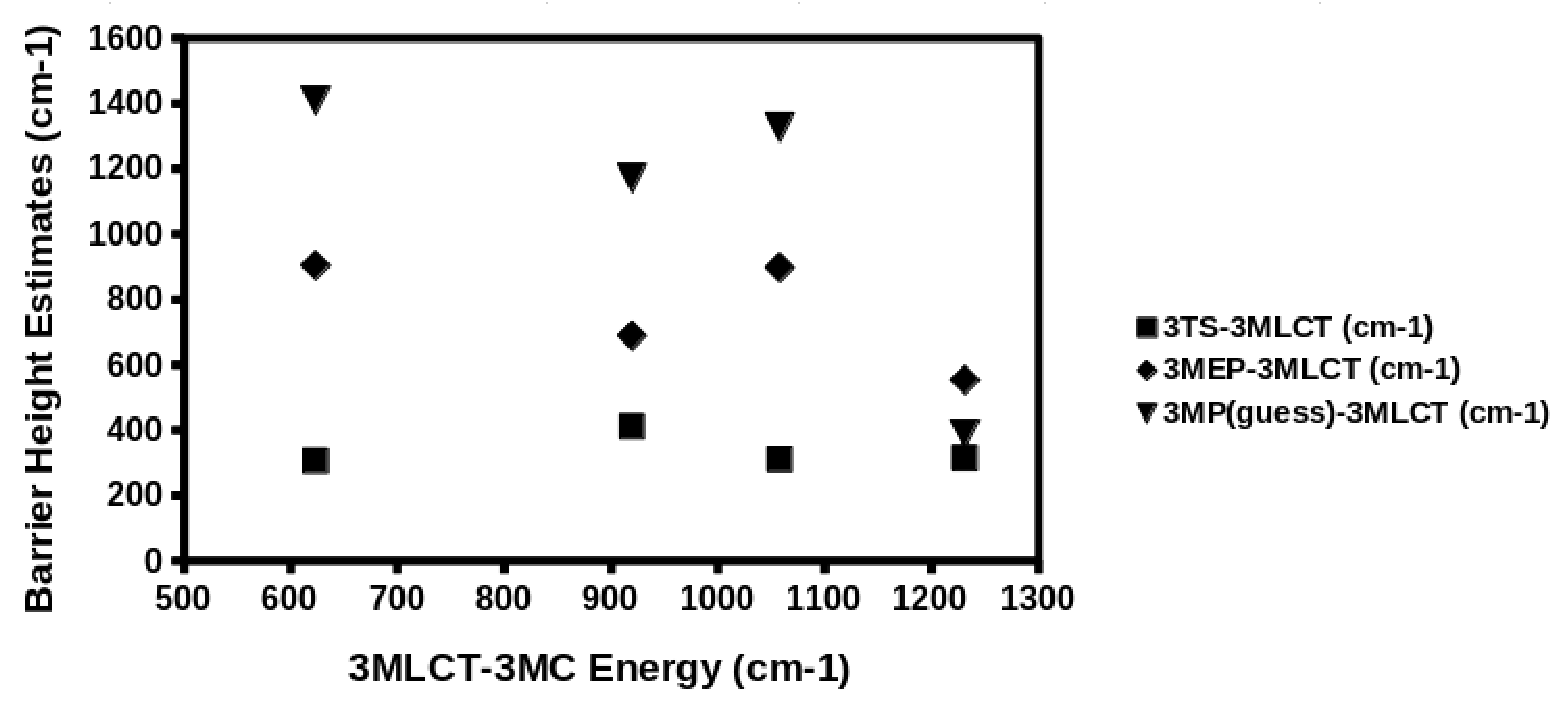}
  \caption{$^3$MLCT $\rightarrow$ $^3$MC barrier heights: black square, 
  accurate TS; black diamond, converged NEB MEP; black downward facing 
  triangle, initial NEB guess MEP.
  \label{fig:BEP2}
  }
\end{figure}

But this derivation was important here because it underlines that we need
a PEC that resembles two intersecting parabolas, which is evidently not the
case in Fig.~\ref{fig:IRC_against_energy_compound_70} because of the plateau
in the IRC curve caused by the presence of a bifurcation.  For this reason,
we do not expect the $^3$TS-$^3$MLCT energy to follow the Bell-Evans-Polanyi
postulate, even approximately.  However the first guess IRC of the NEB does
not go through the bifurcation region and so in in approximate agreement
with the Bell-Evans-Polanyi postulate as shown in 
{\bf Fig.~\ref{fig:BEP2}}.
As might be expected, after a little reflection, the slope of the first
guess NEB MEPs opposite would be opposite to that seen in 
Fig.~\ref{fig:Eave_vs_LI3} for $E_{\mbox{ave}}$.  Also, as previously 
mentioned our best results (namely $^3$TS-$^3$MLCT) are actually 
essentially constant.
% -----------------------------------------------------------------
% -----------------------------------------------------------------

We are now left with a final conundrum.  All our calculations for the
four complexes studied in this article suggest that {\em trans} dissociation
kinetics is actually {\em un}able to explain the observed correlation between
the experimentally-derived $E_{\mbox{ave}}$ and LI3, 
because our theoretically-calculated $^3$TS-$^3$MLCT is complex-independent
to within the expected accuracy of our methodology.
Nevertheless LI3 is positively correlated with the $^3$MLCT-$^3$MC energy 
difference as expected from FMOT and this leads to a strong correlation
between $E_{\mbox{ave}}$ and the $^3$MLCT-$^3$MC energy 
difference $\Delta E^0$.  Why?  Could it be that our 
experimentally-derived $E_{\mbox{ave}}$ reflects something other than 
kinetics?  Afterall, the measured luminescence lifetimes are orders of 
magnitude longer than typical vibrational lifetimes.  This might allow
some sort of quasiequilibrium to take place.
Indeed, our observations up to this point are more consistent with the 
hypothesis that $^3$MLCT states are being removed via a fast 
$^3$MLCT = $^3$MC equilibrium process whose equilibrium constant is 
roughly, 
\begin{equation}
  \frac{[\mbox{$^3$MC}]}{[\mbox{$^3$MLCT}]} 
  = K_{eq} = e^{-\Delta G^0/RT} \propto e^{-\Delta E^0/RT} \, .
  \label{eq:results.8}
\end{equation}
Hence, all other things being equal, a larger $^3$MLCT-$^3$MC means a
more negative value of $\Delta E^0$ and hence a larger ratio of $^3$MC product
in comparison to the $^3$MLCT reactant, effectively removing the luminescent
$^3$MLCT state in favor of the nonluminescent $^3$MC state and leading to
shorter observed luminescence lifetimes which were previously interpretted
in terms of larger values of $E_{\mbox{ave}}$.
Of course, we cannot rule out the presence of other competing pathways
leading to different products and transition states and hence the possibility
of completing kinetics and quasiequilibria.

% \begin{verbatim}
% 
%  +------+
%  | STOP |
%  +------+
% 
% \end{verbatim}

%%%%%%%
% EOF %
%%%%%%%
% ---------------------------------------
\section{Conclusion}
\label{sec:conclusion}
% \input{conclusion.tex}
% ===========================================
% File: conclusion.tex
% Last updated: Wednesday 6 September 2023
% ===========================================

This may be regarded as the third in a series of articles seeking an 
orbital-based explanation of luminescence life-times in ruthenium(II)
polypyridine complexes.  Article I \cite{WJL+14} introduced the use
of the partial density of states (PDOS) for the extraction of a 
ligand-field theory (LFT) like picture from density-functional theory
(DFT) calculations.  Direct visualization of molecular orbitals (MOs) is
insufficient for this purpose because the $e_g^*$ metal MOs mix heavily
with ligand MOs.  However the PDOS analysis allows the assignment of
an energy to these orbitals in a quantitative, albeit fuzzy and both 
basis set and functional dependent, fashion.  This allowed the investigation
of on the order of 100 molecules in Article II \cite{MCA+17} to develop
an orbital-based luminescence index, called LI3, which correlated 
strongly with an average ``activation energy'' $E_{\mbox{ave}}$ calculated
from experimental luminescence lifetimes measured in solution at 
room temperature and liquid nitrogen temperature.  The work in Articles
I and II is based upon the widely assumed $^3$MLCT $\rightarrow$ $^3$MC
{\em trans} distortion mechanism and made use of ideas from frontier
molecular orbital theory (FMOT).  However it was made clear in Article II
that $E_{\mbox{ave}}$ is quite different from an accurately determined
experimental $^3$MLCT $\rightarrow$ $^3$MC activation energy for 
[Ru(bpy)$_3$]$^{2+}$, complex {\bf 6} in the present article, for 
which $E_{\mbox{ave}}$ = 132 cm$^{-1}$ \cite{MCA+17}, but for which experimental
studies in solution gave 3800 cm$^{-1}$ \cite{CM83} and 3960 cm$^{-1}$
depending upon the solvent and details of the analysis.  The present
article corrects this somewhat unsatisfactory state of affairs by carrying
out calculations of the transition states (TSs) and intrinsic reaction
coordinates (IRCs) for four closely related ruthenium(II) bipyridine 
complexes under the same conditions as used to calculate LI3 in Article II,
namely gas phase calculations with the B3LYP functional and the 
6-31G+LANLDZ(Ru) basis set.  Our calculated $^3$MLCT $\rightarrow$ $^3$MC
TS barrier is 316.04 cm$^{-1}$ which may be compared with the previously
calculated theoretical gas phase value of 700 cm$^{-1}$ \cite{SAH+20}.

It is important to note that we used the same nudged elastic band (NEB)
methodology as in Ref.~\cite{SAH+20}, though the basis set and functional 
differ, but made a more thorough investigation of our four molecules using
more accurate TS and IRC searches.  The result carries some surprises.
For example, our NEB maximum energy point (MEP) gives a TS barrier of
553 cm$^{-1}$ (closer to the 700 cm$^{-1}$ value quoted in Ref.~\cite{SAH+20},
particularly when considering that chemical accuracy is often quoted
as 1 kcal/mol = 350 cm$^{-1}$).  But we realized that the NEB MEP is
not sufficiently accurate for our purposes and so used specific TS optimizers
and proved the correctness of our optimized TS by calculation of the IRC.
Examination of Ru-N bond lengths along the IRC showed some surprising results
which can only be explained by a TS corresponding to one of the ligands
symmetrically lengthening its Ru-N bonds, hence going through a sort of
{\em cis} TS consistent with electron transfer to a single ligand, rather
than symmetrically to all three ligands.  The {\em trans} dissociation
then continues along a ridge and then bifurcates into one of two 
symmetry-equivalent minima.  Once this is taken into account, not only is
a lower TS barrier obtained, but a much richer and more complex description 
is obtained of the {\em trans} dissociation mechanism.

Returning to LI3 and the energetics of {\em trans} dissociation, we find
that the $^3$MLCT-$^3$MC energy correlates linearly with LI3 but that the
$^3$TS-$^3$MLCT is, to within the accuracy of our calculation, molecule
independent.  Attempts to make alternative kinetic arguments, such as by
using the Bell-Evans-Polanyi postulate, only lead to ideas in contradiction
with the observed relation between $E_{\mbox{ave}}$ and LI3.  This suggests
that it is the total energy difference of the reaction, rather than the
barrier height, which determines the luminescence lifetime.  That is,
we are looking at a quasielquilibrium property rather than a kinetic
property.  Given that the measured luminescence lifetimes are on the order 
of $\mu$s which is much several orders of magnitude longer than a typical 
vibrational time, then such an assumption is not unreasonable.  

Our picture of what is going on in our study is limited by it being
a gas-phase study with a particular basis set and a particular functional.
It is also limited by the fact that only one luminescence quenching
mechanism has been studied in a family of molecules which is rich in
degrees of freedom and hence also rich in dissociation mechanisms as
shown by the brief review of related work given in the introduction of
this article.  Nevertheless we think that our work brings a new and more
detailed picture of part of what could be going on in the luminescence 
mechanism for this important family of molecules.

%%%%%%%
% EOF %
%%%%%%%
% ---------------------------------------
\section*{Acknowledgement}
\label{sec:thanks}
% \input{thanks.tex}
% ==============================================
% File: thanks.hxv
% Last update: 9 August 2023
% ==============================================

DM and MEC gratefully acknowledge helpful funding from the 
African School on Electronic Structure Methods and Applications (ASESMA),
ASESMANet, and the US-Africa Initiative. 
We thank the following people for insightful discussions: Ragnar Bj\"ornsson,
Isabelle M.\ Dixon, Jean Louis Heully, Valid Mwatati, Max Daku Lat\'evi Lawson, Sandeep Sharma.  In particular, Isabelle M.\ Dixon is gratefully
acknowledged for sharing unpublished computational results with us early
on in our project.  We would like to thank Pierre Girard and
S\'ebastien Morin for technical support in the context
of the Grenoble {\em Centre d'Experimentation du Calcul Intensif en Chimie}
({\em CECIC}) computers used for the calculations reported here. 

%%%%%%
% EOF
%%%%%%
% ---------------------------------------
\newpage
\section*{Supplementary Information}
\label{sec:SI}
% \input{SI.tex}
% ======================================
% File: SI.tex
% Last modified: 9 August 2023
% ======================================
Only plots for complex {\bf 70} have been used in the main article.
The corresponding plots for complexes {\bf 6}, {\bf 73}, and {\bf 74}
are available on-line: 
\begin{enumerate}
  \item Contour plots
  \item NEB optimizations
  \item Converged NEB Minimum Energy Paths
  \item IRC Energy Profiles
  \item Variation of Metal-Ligand Bond Lengths Along the NEB Reaction Path
  \item Variation of Metal-Ligand Bond Lengths Along the IRC
  \item Author contributions
\end{enumerate}

%%%%%%%
% EOF %
%%%%%%%
% ============================================================
\bibliographystyle{myaip}
\bibliography{refs}

\begin{thebibliography}{10}

\bibitem{MMAC20}
D.~Magero, T.~Mestiri, K.~Alimi, and M.~E. Casida,
\newblock {\sf \color{blue}Computational studies of ruthenium and iridium
  complexes for energy sciences and progress on greener alternatives},
\newblock in {\em Green Chemistry and Computational Chemistry: Shared Lessons
  in Sustainability}, edited by L.~Mammino, pages 115--145, Elsevier, 2020,
\newblock preprint: https://arxiv.org/abs/2004.03345.

\bibitem{SCC+94}
J.~P. Sauvage, J.~Collin, J.~C. Chambron, S.~Guillerez, and C.~Coudret,
\newblock {\sf \color{blue}{R}uthenium({II}) and osmium({II}) bis(terpyridine)
  complexes in covalently-linked multicomponent systems: {S}ynthesis,
  electrochemical behavior, absorption spectra, and photochemical and
  photophysical properties},
\newblock Chem. Rev. {\bf 94}, 993 (1994).

\bibitem{N82}
K.~Nakamaru,
\newblock {\sf \color{blue}Synthesis, luminescence quantum yields, and
  lifetimes of trischelated ruthenium ({II}) mixed-ligand complexes including
  3,3'-dimethyl-2,2'-bipyridyl},
\newblock Bull. Chem. Soc. Jpn. {\bf 55}, 2697 (1982).

\bibitem{LKW99}
G.~Liebsch, I.~Klimant, and O.~S. Wolfbeis,
\newblock {\sf \color{blue}Luminescence lifetime temperature sensing based on
  sol-gels and poly(acrylonitrile)s dyed with ruthenium metal-ligand
  complexes},
\newblock Adv. Mat. {\bf 11}, 1296 (1999).

\bibitem{BJ01}
V.~Balzani and A.~Juris,
\newblock {\sf \color{blue}Photochemistry and photophysics of {Ru (II)}
  polypyridine complexes in the bologna group. {F}rom early studies to recent
  developments},
\newblock Coordin. Chem. Rev. {\bf 211}, 97 (2001).

\bibitem{DTL+03}
M.~Duati, S.~Tasca, F.~C. Lynch, H.~Bohlen, J.~G. Vos, S.~Stagni, and M.~D.
  Ward,
\newblock {\sf \color{blue}Enhancement of luminescence lifetimes of mononuclear
  ruthenium({II})-terpyridine complexes by manipulation of the $\sigma$-donor
  strength of ligands},
\newblock J. Inorg. Chem. {\bf 42}, 8377 (2003).

\bibitem{HKZ03}
A.~Harriman, A.~Khatyr, and R.~Ziessel,
\newblock {\sf \color{blue}Extending the luminescence lifetime of
  ruthenium({II}) poly(pyridine) complexes in solution at ambient temperature},
\newblock Dalton Trans. {\bf 10}, 2061 (2003).

\bibitem{MH05}
E.~A. Medlycott and G.~S. Hanan,
\newblock {\sf \color{blue}Designing tridentate ligands for ruthenium({II})
  complexes with prolonged room temperature luminescence lifetimes},
\newblock Chem. Soc. Rev. {\bf 34}, 133 (2005).

\bibitem{NSF+08}
L.~J. Nurkkala, R.~O. Steen, H.~K.~J. Friberg, J.~A. H\"aggstr\"om, P.~V.
  Bernhardt, M.~J. Riley, and S.~J. Dunne,
\newblock {\sf \color{blue}The effects of pendant vs.\ fused thiophene
  attachment upon the luminescence lifetimes and electrochemistry of
  tris(2,2'-bipyridine)ruthenium({II}) complexes},
\newblock Eur. J. Inorg. Chem. {\bf 26}, 4101 (2008).

\bibitem{JWW+10}
S.~Ji, W.~Wu, W.~Wu, P.~Song, K.~Han, Z.~Wang, S.~Liu, H.~Guo, and J.~Zhao,
\newblock {\sf \color{blue}Tuning the luminescence lifetimes of ruthenium({II})
  polypyridine complexes and its application in luminescent oxygen sensing},
\newblock J. Mater. Chem. {\bf 20}, 1953 (2010).

\bibitem{SAH+20}
A.~Soupart, F.~Alary, J.~Heully, P.~I.~P. Elliot, and I.~M. Dixon,
\newblock {\sf \color{blue}Theoretical study of the full photosolvolysis
  mechanism of {[Ru(bpy)$_3$]$^{2+}$}: {P}roviding a general mechanistic
  roadmap for the photochemistry of {[Ru(N$\wedge$N)$_3$]$^{2+}$}-type
  complexes toward both cis and trans photoproducts},
\newblock Inorg. Chem. {\bf 59}, 14679 (2020).

\bibitem{BCCV21}
V.~Balzani, P.~Ceroni, A.~Credi, and M.~Venturi,
\newblock {\sf \color{blue}Ruthenium tris(bipyridine) complexes: {I}nterchange
  between photons and electrons in molecular-scale devices and machines},
\newblock Coord. Chem. Rev. {\bf 433}, 213758 (2021).

\bibitem{WJL+14}
C.~M. Wawire, D.~Jouvenot, F.~Loiseau, P.~Baudin, S.~Liatard, L.~Njenga,
  G.~Kamau, and M.~E. Casida,
\newblock {\sf \color{blue}Density-functional study of luminescence in
  polypyrine ruthenium complexes},
\newblock J. Photochem. and Photobiol. A {\bf 276}, 8 (2014).

\bibitem{MCA+17}
D.~Magero, M.~E. Casida, G.~Amolo, N.~Makau, and L.~Kituyi,
\newblock {\sf \color{blue}Partial density of states ligand field theory
  ({PDOS-LFT}): {R}ecovering a {LFT}-like picture and application to
  photoproperties of ruthenium({II}) polypyridine complexes},
\newblock J. Photochem. Photobiol. A {\bf 348}, 305 (2017).

\bibitem{VC83}
L.~G. Vanquickenborne and A.~Ceulemans,
\newblock {\sf \color{blue}Ligand-field models and the photochemistry of
  coordination compounds},
\newblock Coord. Chem. Rev. {\bf 48}, 157 (1983).

\bibitem{CM83}
J.~V. Caspar and T.~J. Meyer,
\newblock {\sf \color{blue}Photochemistry of {\em tris}(2,2'-bipyridine)
  ruthenium(ii) ion ({[Ru(bpy)$_3$]$^{2+}$}). {S}olvent effects.},
\newblock J. Am. Chem. Soc. {\bf 105}, 5583 (1983).

\bibitem{B01}
I.~B. Bersuker,
\newblock {\sf \color{blue}Modern aspects of the jahn-teller effect theory and
  applications to molecular problems},
\newblock Chem. Rev. {\bf 101}, 1067 (2001).

\bibitem{KZD+06}
I.~Krivokapic, M.~Zerara, M.~{Lawson~Daku}, A.~Vargas, C.~Enachescu, C.~Ambrus,
  P.~{Tregenna-Piggott}, N.~Amstutz, E.~Krausz, and A.~Hauser,
\newblock {\sf \color{blue}Spin-crossover in cobalt({II}) imine complexes},
\newblock Coord. Chem. Rev. {\bf 251}, 364 (2007).

\bibitem{AJ18}
D.~C. Ashley and E.~Jakubikova,
\newblock {\sf \color{blue}Ray-{D}utt and {B}ailar twists in {Fe(II)}-{\em
  tris}(2,2'-bipyridine): Spin states, sterics, and {Fe-N} bond strengths},
\newblock Inorg. Chem. {\bf 57}, 5585 (2018).

\bibitem{VD00}
A.~Vaidyalingam and P.~K. Dutta,
\newblock {\sf \color{blue}Analysis of the photodecomposition products of
  {Ru(bpy)$_3^{2+}$} in various buffers and upon zeolite encapsulation},
\newblock Anal. Chem. {\bf 72}, 5219 (2000).

\bibitem{YYS+15}
N.~Yoshikawa, S.~Yamabe, S.~Sakaki, N.~Kanehisa, T.~Inoue, and H.~Takashima,
\newblock {\sf \color{blue}Transition states of the {$^3$MLCT} to {$^3$MC}
  conversion in {Ru(bpy)$_2$(phen derivative)$^{2+}$} complexes},
\newblock J. Mol. Struct. {\bf 1094}, 98 (2015).

\bibitem{ZBP16}
X.~W. Zhou, P.~L. Burn, and B.~J. Powell,
\newblock {\sf \color{blue}Bond fission and non-radiative decay in
  iridium({III}) complexes},
\newblock Inorg. Chem. {\bf 55}, 5266 (2016).

\bibitem{ZBP17}
X.~W. Zhou, P.~L. Burn, and B.~J. Powell,
\newblock {\sf \color{blue}Correction to ``bond fission and non-radiative decay
  in iridium({III}) complexes''},
\newblock Inorg. Chem. {\bf 56}, 7574 (2017).

\bibitem{SDV+17}
Q.~Sun, B.~Dereka, E.~Vauthey, L.~M.~L. Daku, and A.~Hauser,
\newblock {\sf \color{blue}Ultrafast transient {IR} spectroscopy and {DFT}
  calculations of ruthenium({II}) polypyridyl complexes},
\newblock Chem. Sci. {\bf 8}, 223 (2017).

\bibitem{SDAH18}
A.~Soupart, I.~M. Dixon, F.~Alary, and J.~Heully,
\newblock {\sf \color{blue}{DFT} rationalization of the room-temperature
  luminescence properties of {R}u(bpy)$_3^{2+}$ and {R}u(tpy)$_3^{2+}$:
  $^3${MLCT}-$^3${MC} minimum energy path from neb calculations and emission
  spectra from vres calculations},
\newblock Theor. Chem. Acc. {\bf 137}, 37 (2018).

\bibitem{SAH+18}
A.~Soupart, F.~Alary, J.~Heully, P.~I.~P. Elliott, and I.~M. Dixon,
\newblock {\sf \color{blue}Exploration of the uncharted {$^3$PES} territory for
  {[Ru(bpy)$_3$]$^{2+}$]} {A} new {$^3$MC} minimum prone to ligand loss
  photochemistry},
\newblock Inorg. Chem. {\bf 57}, 3192 (2018).

\bibitem{FHG+19}
M.~Fumanal, Y.~Harabuchi, E.~Gindensperger, S.~Maeda, and C.~Daniel,
\newblock {\sf \color{blue}Excited-state reactivity of
  {[Mn(im)(CO)$_3$(phen)]$^+$}: {A} structural exploration},
\newblock J. Comput. Chem. {\bf 40}, 72 (2019).

\bibitem{AHBV07}
F.~Alary, J.~L. Heully, L.~Bijeire, and P.~Vicendo,
\newblock {\sf \color{blue}Is the {$^3$MLCT} the only photoreactive state of
  polypyridyl complexes?},
\newblock Inorg. chem. {\bf 46}, 3154 (2007).

\bibitem{HAB09}
J.~Heully, F.~Alary, and M.~{Boggio-Pasqua},
\newblock {\sf \color{blue}Spin-orbit effects on the photophysical properties
  of {[Ru(bpy)$_3$]$^{2+}$}},
\newblock J. Chem. Phys. {\bf 131}, 184308 (2009).

\bibitem{DHAE17}
I.~M. Dixon, J.~Heully, F.~Alary, and P.~I.~P. Elliott,
\newblock {\sf \color{blue}Theoretical illumination of highly original
  photoreactive {$^3$MC} states and the mechanism of the photochemistry of
  {Ru(II)} tris(bidentate) complexes},
\newblock Phys. Chem. Chem. Phys. {\bf 19}, 27765 (2017).

\bibitem{M80}
K.~M\"uller,
\newblock {\sf \color{blue}Reaction paths on multidimensional energy
  hypersurfaces},
\newblock Angew. Chem. Int. Ed. Engl. {\bf 19}, 1 (1980).

\bibitem{JMJ98}
H.~J{\'o}nsson, G.~Mills, and K.~W. Jacobsen,
\newblock {\sf \color{blue}Nudged elastic band method for finding minimum
  energy paths of transitions},
\newblock in {\em Classical and quantum dynamics in condensed phase
  simulations}, pages 385--404, World Scientific, 1998.

\bibitem{ABB+21}
V.~{\'A}sgeirsson, B.~O. Birgisson, R.~Bjornsson, U.~Becker, F.~Neese,
  C.~Riplinger, and H.~J{\'o}nsson,
\newblock {\sf \color{blue}Nudged elastic band method for molecular reactions
  using energy-weighted springs combined with eigenvector following},
\newblock Journal of chemical theory and computation {\bf 17}, 4929 (2021).

\bibitem{S03}
H.~B. Schlegel,
\newblock {\sf \color{blue}Exploring potential energy surfaces for chemical
  reactions: {A}n overview of some practical methods},
\newblock J. Comput. Chem. {\bf 24}, 1514 (2003).

\bibitem{S11}
H.~B. Schlegel,
\newblock {\sf \color{blue}Geometry optimization},
\newblock WIREs Comput. Mol. Sci. {\bf 1}, 790 (2011).

\bibitem{PS93}
C.~Peng and H.~B. Schlegel,
\newblock {\sf \color{blue}Combining synchronous transit and quasi-{N}ewton
  methods to find transition states},
\newblock Israel J. Chem. {\bf 33}, 449 (1993).

\bibitem{PASF96}
C.~Peng, P.~Y. Ayala, H.~B. Schlegel, and M.~J. Frisch,
\newblock {\sf \color{blue}Using redundant internal coordinates to optimize
  equilibrium geometries and transition states},
\newblock J. Comput. Chem. {\bf 17}, 49 (1996).

\bibitem{HBK05}
A.~Heyden, A.~T. Bell, and F.~J. Keil,
\newblock {\sf \color{blue}Efficient methods for finding transition states in
  chemical reactions: {C}omparison of improved dimer method and partitionned
  rational function optimization method},
\newblock J. Chem. Phys. {\bf 123}, 224101 (2005).

\bibitem{SZBH12}
S.~M. Sharada, P.~M. Zimmerman, A.~T. Bell, and M.~{Head-Gordon},
\newblock {\sf \color{blue}Automated transition state searches without
  evaluating the {H}essian},
\newblock J. Chem. Theory Comput. {\bf 8}, 5166 (2012).

\bibitem{VR18}
A.~C. Vaucher and M.~Reiher,
\newblock {\sf \color{blue}Minimum energy paths and transition states by curve
  optimization},
\newblock J. Chem. Theory Comput. {\bf 14}, 3091 (2018).

\bibitem{MHO+15}
S.~Maeda, Y.~Harabuchi, Y.~Ono, T.~Taketsugu, and K.~Morokuma,
\newblock {\sf \color{blue}Intrinsic reaction coordinate: {C}alculation,
  bifurcation, and automated search},
\newblock Int. J. Quant. Chem. {\bf 115}, 258 (2015).

\bibitem{EWI+08}
D.~H. Ess, S.~E. Wheeler, R.~G. Iafe, L.~Xu, N.~Celebi-Oelcuem, and K.~N. Houk,
\newblock {\sf \color{blue}Bifurcations on potential energy surfaces of organic
  reactions},
\newblock Angewandte Chemie International Edition {\bf 47}, 7592 (2008).

\bibitem{g09}
M.~J. Frisch et~al.,
\newblock {\sf \color{blue}Gaussian 09 {R}evision {D}.01},
\newblock {G}aussian Inc. Wallingford CT 2009.

\bibitem{NWB+20}
F.~Neese, F.~Wennmohs, U.~Becker, and C.~Riplinger,
\newblock {\sf \color{blue}The orca quantum chemistry program package},
\newblock The Journal of chemical physics {\bf 152}, 224108 (2020).

\bibitem{DHP71}
R.~Ditchfield, W.~Hehre, and J.~A. Pople,
\newblock {\sf \color{blue}Self-consistent molecular-orbital methods. ix. an
  extended gaussian-type basis for molecular-orbital studies of organic
  molecules},
\newblock J. Chem. Phys. {\bf 54}, 724 (1971).

\bibitem{HDP72}
W.~J. Hehre, R.~Ditchfield, and J.~A. Pople,
\newblock {\sf \color{blue}Self—consistent molecular orbital methods. xii.
  further extensions of gaussian—type basis sets for use in molecular orbital
  studies of organic molecules},
\newblock J. Chem. Phys. {\bf 56}, 2257 (1972).

\bibitem{HW85b}
P.~J. Hay and W.~R. Wadt,
\newblock {\sf \color{blue}{\color{blue} \sf Ab initio effective core
  potentials for molecular calculations. Potentials for K to Au including the
  outermost core orbitals}},
\newblock J. Chem. Phys. {\bf 82}, 299 (1985),
\newblock {\color{blue} \sf Ab initio effective core potentials for molecular
  calculations. Potentials for K to Au including the outermost core orbitals}.

\bibitem{SDCF94}
P.~J. Stephens, F.~J. Devlin, C.~F. Chabalowski, and M.~J. Frisch,
\newblock {\sf \color{blue}Ab initio calculation of vibrational absorption and
  circular dichroism spectra using density functional force fields},
\newblock J. Phys. Chem. {\bf 98}, 11623 (1994).

\bibitem{VWN80}
S.~H. Vosko, L.~Wilk, and M.~Nusair,
\newblock {\sf \color{blue}Accurate spin-dependent electron liquid correlation
  energies for local spin density calculations: a critical analysis},
\newblock Can. J. Phys. {\bf 58}, 1200 (1980).

\bibitem{M56}
R.~A. Marcus,
\newblock {\sf \color{blue}On the theory of oxidation-reduction reactions
  involving electron transfer. {I}},
\newblock J. Chem. Phys. {\bf 24}, 966 (1956).

\bibitem{KB99}
F.~Kanoufi and A.~J. Bard,
\newblock {\sf \color{blue}Electrogenerated chemiluminescence. 65. {A}n
  investigation of the oxidation of oxalate by tris(polypyridine) ruthenium
  complexes and the effect of the electrochemical steps on the emission
  intensity},
\newblock J. Chem. Phys. B {\bf 103}, 10469 (1999).

\bibitem{Bertsch}
G.~F. Bertsch,
\newblock {\sf \color{blue}Derivations of {M}arcus's formula},
\newblock http://www.int.washington.edu/users/bertsch/marcus.1.pdf, last
  accessed 6 May 2020.

\bibitem{LK83}
K.~J. Laldler and M.~C. King,
\newblock {\sf \color{blue}The development of transition-state theory},
\newblock J. Phys. Chem. {\bf 87}, 2657 (1983).

\bibitem{BMR96}
P.~F. Barbara, T.~J. Meyer, and M.~A. Ratner,
\newblock {\sf \color{blue}Contemporary issues in electron transfer research},
\newblock J. Phys. Chem. {\bf 100}, 13148 (1996).

\bibitem{B36}
R.~P. Bell,
\newblock {\sf \color{blue}The theory of reactions involving proton transfers},
\newblock Proc. R. Soc. London, Ser. A {\bf 154}, 414 (1936).

\bibitem{EP36}
M.~G. Evans and M.~Polanyi,
\newblock {\sf \color{blue}Further considerations on the thermodynamics of
  chemical equilibria and reaction rates},
\newblock J. Chem. Soc., Faraday Trans. {\bf 32}, 1333 (1936).

\end{thebibliography}
% % ----------------------------------------------------------
% \appendix
% \section{Supplementary Information}
% \label{sec:suppinfo}
% \input{supp.tex}
% % ----------------------------------------------------------
\end{document}